\documentclass[9pt]{article}
\usepackage{amsmath,amssymb}
\usepackage{indentfirst}

\begin{document}

\title{The Deformation Quantization of the Scalar Fields}
\author{Jie Wu \thanks{Basic Courses Department, Sion-German University of Applied Sciences,No.2 Yashen Road, Jinnan District, Tianjin, P.R.China,
 email:ewujie@sina.com}
       \and
       Mai Zhou \thanks{College of Mathematical Sciences, Nankai University,
       No.94 Weijin Road, Nankai District, Tianjin, P.R.China, email:zhoumai@nankai.edu.cn}}

\maketitle

\begin{abstract}
In this paper the deformation quantization is constructed in the case of scalar fields on Minkowski space-time based on the equal time Poisson brackets. We construct the star products at three levels concerning fields, Hamiltonian functionals and their underlying structure called Hamiltonian functions in a special sense. Which means the star products of fields, functionals, Hamiltonian functions, and ones between the fields and functionals. As bases of star products the Poisson brackets at different levels are generalized, constructed and discussed in a systematical way, where the Poisson brackets like canonical and time-equal ones. For both of the star products and Poisson brackets the construction at level of Hamiltonian functions plays the essential role. Actually, the discussion for the case of Hamiltonian functions includes the key information about Poisson brackets and the star products in our approach. All of other Poisson brackets and star products in this paper are based on ones of Hamiltonian functions, and the Poisson brackets and the star products at three level are compatible. To discuss the Poisson brackets and the star products in the case of scalar fields, we introduce the notion of algebra of Euler-Lagrange operators which related to variation calculus closely.
\end{abstract}

\newpage
\tableofcontents

\newpage
\section{Introduction}

It is well known that the deformation quantization means constructing non-commutative multiplication as a deformation of commutative one carried over an algebra at classical level such that the semi-classical limit of non-commutative multiplication is Poisson bracket. In the present paper we discuss the deformation quantization in the case of scalar fields, which is the infinite dimensional generalization of deformation quantization on finite dimensional phase space. The deformation quantization of the classical fields theory should be worthy to discuss the problems in quantum fields theory, for example, the perturbative algebraic quantum field theory(pAQFT). In the framework of pAQFT, a crucial stage is construction of star products on a Lorenz manifold starting from Peierls bracket concerning Lagrangian formulation of classical fields theory(Klaus Fredenhagen, Katarzyna Rejzner\cite{a},\cite{b} and their references).

In order to explain our idea we start from a simple example. Now we recall simply some background of physics in the case of real scalar fields at classical level. In the sense of Lagrangian formulation a real scalar field is a real function on Minkowski space-time $\varphi(t,x)$, $(t,x)\in\mathbb{R}_{t}\bigoplus\mathbb{R}_{x}^{3}$, a Lagrangian density arising from some physical system is a function with form $ \mathcal{L}(\varphi,\dot{\varphi},\nabla_{x}\varphi)$, here $\dot{\varphi}=\frac{\partial\varphi}{\partial t}$, $\nabla_{x}\varphi$ is gradient
of $\varphi$ about space variable $x$. The Lagrangian is an integral of Lagrangian density about space variables $\int\mathcal{L}(\varphi,\dot{\varphi},\nabla_{x}\varphi)dx$. The conjugate field of $\varphi(t,x)$ is defined by $\pi(t,x)=\frac{\partial\mathcal{L}}{\partial\dot{\varphi}}$, Hamitonian density and Hamitonian are $\mathcal{H}(\varphi,\pi,\nabla_{x}\varphi)=\pi\dot{\varphi}-\mathcal{L}$, and
$$H(\varphi,\pi)=\int\mathcal{H}(\varphi,\pi,\nabla_{x}\varphi)dx$$
respectively. Usually the Poisson bracket of $\pi(t,x),\varphi(t,x)$ is defined in the following way
$$\{\pi(t,x),\varphi(t,y)\}=-i\delta(x-y),\{\pi(t,x),\pi(t,y)\}=0,\{\varphi(t,x),\varphi(t,y)\}=0.$$
Where $\delta(\cdot)$ is Dirac function on $\mathbb{R}_{x}^{3}$.

There are three objects at three levels concerning functional $H(\varphi,\pi)$ shown in the following diagram
$$\mathcal{H}(u,v,\xi)\dashleftarrow\dashrightarrow\textit{level of functions}$$
$$\mathcal{H}(\pi(t,x),\varphi(t,x),\nabla_{x}\varphi(t,x))\dashleftarrow\dashrightarrow\textit{level of fields}$$
$$H(\varphi,\pi)\dashleftarrow\dashrightarrow\textit{level of functionals}$$
Where $u,v\in\mathbb{R},\xi\in\mathbb{R}^{3}$. Our purpose is to construct the star products at levels of fields and functionals, or, star products between the fields and functionals. But we found that in principle the operation of star products occurs at level of functions shown in above diagram. In another word the case at level of functions includes some key points of the star products. Thus we need to construct the star products at level of functions firstly. It is well known that the Poisson bracket is start point of deformation quantization, thus, to construct the star products at all of three level mentioned above, we generalize the Poisson brackets to more general forms and discuss that systematically at three level. We will show that the star products over here are natural generalization of Weyl star product to the case of infinite dimension. Actually, we introduce the notions of Euler-Lagrange derivatives which play the roles of partial derivatives on finite
dimensional phase space in our setting, such that the forms of the star products in this paper like the classical Weyl star product very much.
In this paper the Hamiltonian functionals are type of local functionals in the sense of \cite{a}, actually, it seems that the constructions in the present paper working on instantaneous phase space under Hamiltonian formalism play the role
of Cauchy data for the covariant framework in Klaus Fredenhagen, Katarzyna Rejzner\cite{a},\cite{b}.

This paper is organized as following. From section2 to section4 we discuss the deformation quantization for the case of real scalar fields. In section2 we discuss the notion of Euler-Lagrange operators connecting the variation of functionals closely and playing the important roles in  discussion of the star products. In section3 the Poisson brackets and star products is discussed at level of functions in the sense of above diagram. In section4 we discuss the star products at levels of fields and functionals, furthermore, star products between the fields and functionals. In section5 we generalize the discussion to the case of complex scalar fields.

\section{The Euler-Lagrange operators}
\newtheorem{theorem}{Theorem}[section]
\newtheorem{definition}{Definition}[section]
\newtheorem{propsition}{Propsition}[section]
\newtheorem{remark}{Remark}[section]
\numberwithin{equation}{section}

\subsection{The basic notations}

To construct the deformation quantization starting from Hamitonian formulation of the classical fields theory mentioned above we need to generalize that.  At beginning we introduce the notion of Euler-Lagrange operators which will play key roles in the discussion about Poisson brackets and deformation quantization. We consider the set of variables
$$\{(u_{\alpha},\xi_{\beta})|u_{\alpha},\xi_{\beta}\in\mathbb{R},\alpha,\beta\in\mathbb{N}^{n} \}$$
where $\mathbb{N}^{n}$ is the set of multiple indices $\alpha=(\alpha_{1},\cdots,\alpha_{n}),(\alpha_{i}\in\mathbb{N},i=1,\cdots,n)$. Let $\mathcal{H}^{\infty}$ denote the set of real smooth functions $f(\cdots,u_{\alpha},\cdots,\xi_{\beta},\cdots)$ called Hamiltonian functions, for simplicity we denote those functions by $f(u_{\alpha},\xi_{\beta})$, where the variables of functions is a finite subset of
$\{u_{\alpha},\xi_{\beta}\}_{\alpha,\beta\in\mathbb{N}^{n}}$. For a Hamiltonian function
$f(u_{\alpha},\xi_{\beta})\in\mathcal{H}^{\infty}$, we define the
Hamiltonian density in the following way
$$f(\partial^{\alpha}_{x}\varphi(x),\partial^{\beta}_{x}\pi(x))=f(u_{\alpha},\xi_{\beta})|_{u_{\alpha}=\partial^{\alpha}_{x}\varphi(x),\xi_{\beta}=\partial^{\beta}_{x}\pi(x)},$$
where $\varphi(x),\pi(x)\in\mathcal{S}(\mathbb{R}^{n})\cap\mathrm{C}^{\infty}_{R}(\mathbb{R}^{n})$, $\mathcal{S}(\mathbb{R}^{n})$ is Schwartz space on
$\mathbb{R}^{n}$, $\mathrm{C}^{\infty}_{R}(\mathbb{R}^{n})$ denotes the space of real values smooth functions. Let $\mathcal{H}_{den}$ denote the set of Hamiltonian densities. It is easy to check that both of $\mathcal{H}^{\infty}$ and $\mathcal{H}_{den}$ are algebras. Furthermore, we consider the functionals of Hamiltonian type with following form
$$F(\varphi,\pi)=\int_{\mathbb{R}^{n}}f(\partial^{\alpha}_{x}\varphi(x),\partial^{\beta}_{x}\pi(x))dx,$$
where $f(u_{\alpha},\xi_{\beta})\in\mathcal{H}^{\infty},\varphi(x),\pi(x)\in\mathcal{S}(\mathbb{R}^{n})\cap\mathrm{C}^{\infty}_{R}(\mathbb{R}^{n})$.
We call the above functional the Hamiltonian functional. Due to the reason of above integral being well defined, we assume that the Hamiltonian function in the integral satisfies the following condition
\begin{equation}
 f(u_{\alpha},\xi_{\beta})|_{u_{\alpha}=\xi_{\beta}=0}=0
\end{equation}
We call expression (2.1) the condition $\mathrm{B}$. Actually, under the condition $\mathrm{B}$ we know that in a neighborhood of origin we have
$$|f(u_{\alpha},\xi_{\beta})|\leq C\sum(|u_{\alpha}|+|\xi_{\beta}|).$$
Thus, it is easy to check that the Hamiltonian density
$f(\partial^{\alpha}_{x}\varphi(x),\partial^{\beta}_{x}\pi(x))\in\mathcal{S}(\mathbb{R}^{n})$
under the condition $\mathrm{B}$. The set of Hamiltonian functionals is denoted by $H^{\infty}$ where the Hamiltonian functions satisfy the condition $\mathrm{B}$.
Our purpose is to construct the deformation quantization on $\mathcal{H}_{den}$ and $H^{\infty}$, but we will work on $\mathcal{H}^{\infty}$ in principle.\\

\subsection{The Euler-Lagrange operators}
Now we introduce the notion of the Euler-Lagrange operators. Firstly, we introduce some formal notations, let
$$\partial_{u,x;\alpha}\doteq\frac{\partial}{\partial u_{\alpha}}\partial_{x}^{\alpha},\;
\partial_{\xi,x;\beta}\doteq\frac{\partial}{\partial\xi_{\beta}}\partial_{x}^{\beta},\;\alpha,\beta\in\mathbb{N}^{n}. $$
We define the multiplication of $\partial_{u,x;\alpha},\partial_{\xi,x;\beta}$ as following

\begin{equation}
\begin{array}{c}
  \partial_{u,x;\alpha_{1}}\circ\cdots\circ\partial_{u,x;\alpha_{k}}
\circ\partial_{\xi,x;\beta_{1}}\circ\cdots\circ\partial_{\xi,x;\beta_{m}}\\
  \doteq\frac{\partial^{k+m}}{\partial u_{\alpha_{1}}\cdots\partial u_{\alpha_{k}}\partial\xi_{\beta_{1}}\cdots\partial\xi_{\beta_{m}}}
\partial_{x}^{\alpha_{1}+\cdots+\alpha_{k}+\beta_{1}+\cdots+\beta_{m}},
\end{array}
\end{equation}
where $\alpha_{1},\cdots,\alpha_{k},\beta_{1},\cdots,\beta_{m}\in\mathbb{N}^{n}$.
Let $\mathcal{L}$ denote the algebra over $\mathbb{R}$ generated by $\{\partial_{u,x;\alpha},\partial_{\xi,x;\beta},1\}_{\alpha,\beta\in\mathbb{N}^{n}}$ with the multiplication defined in (2.2).
It is easy to see that an element in $\mathcal{L}$ is a polynomial of $\partial_{u,x;\alpha},\partial_{\xi,x;\beta}$, we call this
polynomial the Euler-Lagrange operators.

Now we introduce the following space of distributions
$$Delta_{x,y}=C^{\infty}(\mathbb{R}_{x}^{n}\times\mathbb{R}_{y}^{n})\otimes\{\partial_{x}^{\alpha}\delta(x-y),1\}_{\alpha\in\mathbb{N}^{n}}.$$
Then the Euler-Lagrange operators can be consider as the following maps
$$\mathcal{H}^{\infty}\otimes Delta_{x,y}\rightarrow\mathcal{H}^{\infty}\otimes Delta_{x,y},$$
or,
$$\mathcal{H}^{\infty}\otimes C^{\infty}(\mathbb{R}_{x}^{n})\rightarrow\mathcal{H}^{\infty}\otimes C^{\infty}(\mathbb{R}_{x}^{n}).$$
For example, we define

\begin{equation}
\begin{array}{c}
  (\partial_{u,x;\alpha_{1}}\circ\cdots\circ\partial_{u,x;\alpha_{k}}\circ\partial_{\xi,x;\beta_{1}}
\circ\cdots\circ\partial_{\xi,x;\beta_{m}})(f(u_{\alpha},\xi_{\beta})Q(x,y)) \\
  \doteq\frac{\partial^{k+m}f(u_{\alpha},\xi_{\beta})}
{\partial u_{\alpha_{1}}\cdots\partial u_{\alpha_{k}}\partial\xi_{\beta_{1}}\cdots\partial\xi_{\beta_{m}}}
\partial_{x}^{\alpha_{1}+\cdots+\alpha_{k}+\beta_{1}+\cdots+\beta_{m}}Q(x,y),
\end{array}
\end{equation}
generally, $P(\partial_{u,x;\alpha},\partial_{\xi,x;\beta})(f(u_{\alpha},\xi_{\beta})Q(x,y))$ for an Euler-Lagrange operator\\ $P(\partial_{u,x;\alpha},\partial_{\xi,x;\beta})\in\mathcal{L}$. Where $f(u_{\alpha},\xi_{\beta})\in\mathcal{H}^{\infty}$ and
$Q(x,y)\in Delta_{x,y}$.

Furthermore, starting from the Euler-Lagrange operators we can also define the following maps
$$\mathcal{H}_{den}\otimes Delta_{x,y}\rightarrow\mathcal{H}_{den}\otimes Delta_{x,y},$$
and,
$$\mathcal{H}_{den}\otimes C^{\infty}(\mathbb{R}_{x}^{n})\rightarrow\mathcal{H}_{den}\otimes C^{\infty}(\mathbb{R}_{x}^{n}).$$
Actually, we have

\begin{equation}
\begin{array}{c}
  P(\partial_{u,x;\alpha},\partial_{\xi,x;\beta})(f(\partial_{x}^{\alpha}\varphi(x),\partial_{x}^{\beta}\pi(x))Q(x,y)) \\
  \doteq P(\partial_{u,x;\alpha},\partial_{\xi,x;\beta})(f(u_{\alpha},\xi_{\beta})Q(x,y))
  |_{u_{\alpha}=\partial_{x}^{\alpha}\varphi(x),\xi_{\beta}=\partial_{x}^{\beta}\pi(x)},
\end{array}
\end{equation}
where $f(\partial_{x}^{\alpha}\varphi(x),\partial_{x}^{\beta}\pi(x))\in\mathcal{H}_{den}$. If we fix the Hamiltonian density
$f(\partial_{x}^{\alpha}\varphi(x),\partial_{x}^{\beta}\pi(x))$, we get a linear partial differential operator with smooth coefficients
denoted by
$P(\partial_{u,x;\alpha},\partial_{\xi,x;\beta})(f,x)$. That means we have

\begin{equation}
\begin{array}{c}
 P(\partial_{u,x;\alpha},\partial_{\xi,x;\beta})(f,x)Q(x,y) \\
 \doteq P(\partial_{u,x;\alpha},\partial_{\xi,x;\beta})(f(\partial_{x}^{\alpha}\varphi(x),\partial_{x}^{\beta}\pi(x))Q(x,y))
\end{array}
\end{equation}
We call $P(\partial_{u,x;\alpha},\partial_{\xi,x;\beta})(f,x)$ the related Euler-Lagrange operator. From the definition of related
Euler-Lagrange operator (2.5), it is easy to check that we have

\begin{propsition}
For two Euler-Lagrange operators $P_{1}(\partial_{u,x;\alpha},\partial_{\xi,x;\beta})$,\\
$P_{2}(\partial_{u,x;\alpha},\partial_{\xi,x;\beta})$
we have
\begin{equation}
\begin{array}{c}
  [P_{1}(\partial_{u,x;\alpha},\partial_{\xi,x;\beta})\circ P_{2}(\partial_{u,x;\alpha},\partial_{\xi,x;\beta})](f,x)Q(x,y) \\
  =P_{1}(\partial_{u,x;\alpha},\partial_{\xi,x;\beta})[P_{2}(\partial_{u,x;\alpha},\partial_{\xi,x;\beta})(f,x)Q(x,y)].
\end{array}
\end{equation}
\end{propsition}

To discuss deformation quantization we need to introduce the notion of Euler-Lagrange derivatives as following
\begin{equation}
  \partial_{u,x}\doteq\sum_{\alpha\in\mathbb{N}^{n}}\frac{\partial}{\partial u_{\alpha}}\partial_{x}^{\alpha},\;
  \partial_{\xi,x}\doteq\sum_{\beta\in\mathbb{N}^{n}}\frac{\partial}{\partial \xi_{\beta}}\partial_{x}^{\beta}.
\end{equation}
$\partial_{u,x},\partial_{\xi,x}$ are formal infinite sums, however for a Hamiltonian function $f(u_{\alpha},\xi_{\beta})$ or a Hamiltonian density
$f(\partial_{x}^{\alpha}\varphi(x),\partial_{x}^{\beta}\pi(x))$ we know that
$$\partial_{u,x}(f(u_{\alpha},\xi_{\beta})Q(x,y)),\;\partial_{u,x}(f(\partial_{x}^{\alpha}\varphi(x),\partial_{x}^{\beta}\pi(x))Q(x,y))$$
are finite sums. Formally we can define the power of $\partial_{u,x}$ and $\partial_{\xi,x}$ to be
$$\partial_{u,x}^{k}\doteq\overset{k-times}{\overbrace{\partial_{u,x}\circ\cdots\circ\partial_{u,x}}},\;
\partial_{\xi,x}^{k}\doteq\overset{k-times}{\overbrace{\partial_{\xi,x}\circ\cdots\circ\partial_{\xi,x}}}.$$
It is obvious that $\partial_{u,x}^{k}(f,x),\partial_{\xi,x}^{k}(f,x)$ are well defined. Actually, the behaviour of
$\partial_{u,x}^{k}(f,x),\partial_{\xi,x}^{k}(f,x)$ is same as one of related Euler-Lagrange operators.

\subsection{The dual Euler-Lagrange operators }

Similar to what we do in previous subsection we introduce the following notations
$$D_{u,x;\alpha}\doteq (-\partial_{x})^{\alpha}\frac{\partial}{\partial u_{\alpha}},\;
D_{\xi,x;\beta}\doteq (-\partial_{x})^{\beta}\frac{\partial}{\partial\xi_{\beta}},\alpha,\beta,\in\mathbb{N}^{n}£¬$$
where $(-\partial_{x})^{\alpha}\doteq(-\partial_{x_{1}})^{\alpha_{1}}\cdots(-\partial_{x_{n}})^{\alpha_{n}}.$
We define the multiplication as following

\begin{equation}
\begin{array}{c}
  D_{u,x;\alpha_{1}}\circ\cdots\circ D_{u,x;\alpha_{k}}\circ D_{\xi,x;\beta_{1}}\circ\cdots\circ D_{\xi,x;\beta_{m}} \\
  \doteq(-\partial_{x})^{\alpha_{1}+\cdots+\alpha_{k}+\beta_{1}+\cdots+\beta_{m}}
  \frac{\partial^{k+m}}{\partial u_{\alpha_{1}}\cdots\partial u_{\alpha_{k}}\partial\xi_{\beta_{1}}\cdots\partial\xi_{\beta_{m}}},
\end{array}
\end{equation}
where $\alpha_{1},\cdots,\alpha_{k},\beta_{1},\cdots,\beta_{m}\in\mathbb{N}^{n}$. Let $\mathcal{L}^{t}$ be an algebra over $\mathbb{R}$
generated by $\{D_{u,x;\alpha},D_{\xi,x;\beta},1\}_{\alpha,\beta\in\mathbb{N}^{n}}$ with multiplication defined in (2.8),
the elements in $\mathcal{L}^{t}$ are called the dual Euler-Lagrange operators.

The dual Euler-Lagrange operator can be considered as an operator acting on $\mathcal{H}_{den}$, for example, we have

\begin{equation}
\begin{array}{c}
  (D_{u,x;\alpha_{1}}\circ\cdots\circ D_{u,x;\alpha_{k}}\circ D_{\xi,x;\beta_{1}}\circ\cdots\circ D_{\xi,x;\beta_{m}})
f(\partial_{x}^{\alpha}\varphi(x),\partial_{x}^{\beta}\pi(x)) \\
  \doteq(-\partial_{x})^{\alpha_{1}+\cdots+\alpha_{k}+\beta_{1}+\cdots+\beta_{m}}
  [\frac{\partial^{k+m}f(\partial_{x}^{\alpha}\varphi(x),\partial_{x}^{\beta}\pi(x))}{\partial u_{\alpha_{1}}\cdots\partial u_{\alpha_{k}}\partial\xi_{\beta_{1}}\cdots\partial\xi_{\beta_{m}}}],
\end{array}
\end{equation}
where $f(\partial_{x}^{\alpha}\varphi(x),\partial_{x}^{\beta}\pi(x))\in\mathcal{H}_{den}$. Now we have

\begin{propsition}
For a Euler-Lagrange operator $P(\partial_{u,x;\alpha},\partial_{\xi,x;\beta})$ we have
\begin{equation}
\begin{array}{c}
  [P(\partial_{u,x;\alpha},\partial_{\xi,x;\beta})(f,x)]^{t}(1_{x}) \\
   =P(D_{u,x;\alpha},D_{\xi,x;\beta})f(\partial_{x}^{\alpha}\varphi(x),\partial_{x}^{\beta}\pi(x)).
\end{array}
\end{equation}
\end{propsition}

We define the dual Euler-Lagrange derivative to be following formal infinite sums
\begin{equation}
  L_{u,x}\doteq\sum_{\alpha\in\mathbb{N}^{n}}D_{u,x;\alpha},\;
  L_{\xi,x}\doteq\sum_{\beta\in\mathbb{N}^{n}}D_{\xi,x;\beta}.
\end{equation}
Then we have
\begin{propsition}
\begin{equation}
  [\partial_{u,x}^{k}(f,x)]^{t}(1_{x})=L_{u,x}^{k}f(\partial_{x}^{\alpha}\varphi(x),\partial_{x}^{\beta}\pi(x)),
\end{equation}
\begin{equation}
[\partial_{\xi,x}^{k}(f,x)]^{t}(1_{x})=L_{\xi,x}^{k}f(\partial_{x}^{\alpha}\varphi(x),\partial_{x}^{\beta}\pi(x)).
\end{equation}
\end{propsition}
\begin{remark}
In this section the operators or derivative are named by Euler-Lagrange because they concern the operation of variation closely.
In fact we have
\begin{equation}
  \frac{\delta f(\partial_{x}^{\alpha}\varphi(x),\partial_{x}^{\beta}\pi(x))}{\delta\varphi(y)}
  =\partial_{u,x}(f,x)\delta(x-y),
\end{equation}
\begin{equation}
  \frac{\delta f(\partial_{x}^{\alpha}\varphi(x),\partial_{x}^{\beta}\pi(x))}{\delta\pi(y)}
  =\partial_{\xi,x}(f,x)\delta(x-y),
\end{equation}
\begin{equation}
  \frac{\delta F}{\delta\varphi}=L_{u,x}f(\partial_{x}^{\alpha}\varphi(x),\partial_{x}^{\beta}\pi(x)),
\end{equation}
\begin{equation}
  \frac{\delta F}{\delta\pi}=L_{\xi,x}f(\partial_{x}^{\alpha}\varphi(x),\partial_{x}^{\beta}\pi(x)).
\end{equation}
\end{remark}

\section{The star products on $\mathcal{H}^{\infty}$}
\newtheorem{thm}{Theorem}[section]
\newtheorem{defn}{Definition}[section]
\newtheorem{prop}{Propsition}[section]
\newtheorem{rem}{Remark}[section]
\numberwithin{equation}{section}

The discussion about the Euler-Lagrange operators suggests us the operations of star products occur at level of Hamiltonian functions really. Thus we discuss the star products on $\mathcal{H}^{\infty}$ firstly.

\subsection{Definition of the Poisson brackets on $\mathcal{H}^{\infty}$}
\begin{definition}
For $P(x,y)\in Delta_{x,y}$, we call $P(x,y)$ is symmetric, if
\begin{equation}
  <P(x,y),\varphi(x,y)>=<P(x,y),(P^{\ast}_{x,y}\varphi)(x,y)>,
\end{equation}
we call $P(x,y)$ is anti-symmetric, if
\begin{equation}
  <P(x,y),\varphi(x,y)>=-<P(x,y),(P^{\ast}_{x,y}\varphi)(x,y)>,
\end{equation}
where $\varphi(x,y)\in C_{0}^{\infty}(\mathbb{R}_{x}^{n}\times\mathbb{R}_{y}^{n})$, $P_{x,y}$ is permutation map, $P_{x,y}(x,y)=(y,x)$.
\end{definition}
\begin{remark}
Actually, the symmetric condition is equivalent to the following one
$$<P(x,y),\varphi(x)\psi(y)>=<P(x,y),(P^{\ast}_{x,y}\varphi)(x)\psi(y)>,\varphi)(x),\psi(x)\in C_{0}^{\infty}(\mathbb{R}^{n}).$$
Let $$P(x,y)=\sum_{\gamma\in\mathbb{N}^{n}}\phi_{\gamma}(x,y)\partial_{x}^{\gamma}\delta(x-y),$$
then we have $$P(x,y)=\sum_{\gamma\in\mathbb{N}^{n}}(-1)^{|\gamma|}\phi_{\gamma}(x,y)\partial_{y}^{\gamma}\delta(x-y),$$
because $\partial_{x}^{\gamma}\delta(x-y)=(-1)^{|\gamma|}\partial_{y}^{\gamma}\delta(x-y)$.
On the other hand, noting
 $$<P(x,y),\varphi(x)\psi(y)>=\sum_{\gamma\in\mathbb{N}^{n}}\int_{\mathbb{R}^{n}}(-1)^{|\gamma|}\partial_{x}^{\gamma}
(\phi_{\gamma}(x,y)\varphi(x))|_{x=y}\psi(y)dy,$$
and
$$<P(x,y),\varphi(y)\psi(x)>=\sum_{\gamma\in\mathbb{N}^{n}}\int_{\mathbb{R}^{n}}(-1)^{|\gamma|}\partial_{y}^{\gamma}
(\phi_{\gamma}(x,y)\varphi(y))|_{x=y}\psi(x)dx,$$
finally, we know that the symmetric condition is equivalent to the following formula
$$\sum_{\gamma\in\mathbb{N}^{n}}\partial_{y}^{\gamma}[((-1)^{|\gamma|}\phi_{\gamma}(y,x)-\phi_{\gamma}(x,y))\varphi(x)]|_{x=y}=0,$$
where $\varphi(x)\in C_{0}^{\infty}({\mathbb{R}^{n}})$.
The case of anti-symmetric condition is similar.
\end{remark}
\begin{definition}
Let $P(x,y)\in Delta_{x,y}$, for two Hamiltonian functions
$f(u_{\alpha},\xi_{\beta})$, $g(v_{\alpha},\eta_{\beta})$ in $\mathcal{H}^{\infty}$, their Poisson bracket is defined in the following way.
We assign $f(u_{\alpha},\xi_{\beta})$, $g(v_{\alpha},\eta_{\beta})$ to the variables $x,y\in\mathbb{R}^{n}$ respectively.
When $P(x,y)$ is symmetric, we define
\begin{equation}
  \{f(u_{\alpha},\xi_{\beta}),g(v_{\alpha},\eta_{\beta})\}_{P}\doteq(\partial_{u,x}\partial_{\eta,y}-\partial_{\xi,x}\partial_{v,y})
  f(u_{\alpha},\xi_{\beta})g(v_{\alpha},\eta_{\beta})P(x,y).
\end{equation}
When $P(x,y)$ is anti-symmetric, we define
\begin{equation}
  \{f(u_{\alpha},\xi_{\beta}),g(v_{\alpha},\eta_{\beta})\}_{P}\doteq(\partial_{u,x}\partial_{\eta,y}+\partial_{\xi,x}\partial_{v,y})
  f(u_{\alpha},\xi_{\beta})g(v_{\alpha},\eta_{\beta})P(x,y).
\end{equation}
\end{definition}
\begin{remark}
The Poisson brackets in definition3.2 are much different from the case of finite dimensional phase space, in fact, which can be considered as maps
$$\mathcal{H}^{\infty}\times\mathcal{H}^{\infty}\rightarrow\mathcal{H}^{\infty}\otimes\mathcal{H}^{\infty}\otimes Delta_{x,y}.$$
\end{remark}
\begin{remark}
Because our purpose is to discuss the case of fields, for example,
 $f(\partial_{x}^{\alpha}\varphi(x),\partial_{x}^{\beta}\pi(x))$, or functional $F(\varphi,\pi)$, we need to assign the Hamiltonian function
 $f(u_{\alpha},\xi_{\beta})$ to the variable $x\in\mathbb{R}^{n}$ in definition3.2.
\end{remark}
\begin{remark}
It is easy to check that Poisson brackets (3.3), (3.4) are anti-symmetric, bilinear and derivatives for first and second variables respectively. In this sense $\mathcal{H}^{\infty}$ looks like Poisson algebras.
\end{remark}

\subsection{Multiple Poisson brackets and Jacobi identities}
In this subsection we discuss what happen when we take Poisson bracket repeatedly, for example, the Poisson bracket
$\{h,\{f,g\}_{P}\}_{P}$, or, more general \\
$\{h,\{f_{1},\{\cdots,\{f_{k},g\}_{P}\cdots\}_{P}\}_{P}\}_{P}$. For multiple Poisson brackets, for example
$\{h,\{f_{1},\{\cdots,\{f_{k},g\}_{P}\cdots\}_{P}\}_{P}\}_{P}$, we will work on tenser space which looks like
$$\mathcal{H}^{\infty}\otimes\cdots\otimes\mathcal{H}^{\infty}\otimes Delta_{z_{1},x}\otimes\cdots\otimes Delta_{z_{k},x},z_{i}\neq z_{j},i\neq j.$$
Thus the multiplications of distributions, for example $\delta(x-z_{1})$, $\delta(x-z_{2})$, $\cdots$, $\delta(x-z_{k})$
or their derivatives, will appear. Here we consider $\delta(x-z_{1})$, $\delta(x-z_{2})$, $\cdots$, $\delta(x-z_{k})$ as distributions on
$\mathbb{R}_{x}^{n}\oplus\mathbb{R}_{z_{1}}^{n}\oplus\cdots\mathbb{R}_{z_{k}}^{n}$, actually they are oscillatory integrals on
$\mathbb{R}_{x}^{n}\oplus\mathbb{R}_{z_{1}}^{n}\oplus\cdots\mathbb{R}_{z_{k}}^{n}$ with wave front sets as followings
(Lars H\"{o}rmander \cite{e} Theorem8.1.9)
$$WF\delta(x-z_{i})=\{(x,\cdots,z_{i},x,z_{i+1},\cdots;\xi,0,\cdots,0,-\xi,0,\cdots,0)|\xi\in\mathbb{R}^{n}\setminus\{0\}\}.$$
Thus multiplications of $\delta(x-z_{1})$, $\delta(x-z_{2})$, $\cdots$, $\delta(x-z_{k})$ are well defined and commutative
(Lars H\"{o}rmander \cite{e} Theorem8.2.10).

Furthermore, we consider the partial Euler-Lagrange operators $\partial_{u,x}^{(i,j)}$, $\partial_{\xi,x}^{(i,j)}$
acting on the tenser space
$$\mathcal{H}^{\infty}\otimes\cdots\otimes\overset{i-th}{\overbrace{\mathcal{H}^{\infty}}}\otimes\cdots\otimes\mathcal{H}^{\infty}\otimes
  \cdots\otimes\overset{j-th}{\overbrace{Delta_{x,y}}}\otimes\cdots.$$
Here the first index $i$ corresponds to the position of factor with type of $\mathcal{H}^{\infty}$ and $i-th$ factor $\mathcal{H}^{\infty}$ is assigned to the variable $x$. The second index $j$ corresponds to the position of factor with type of $Delta_{x,y}$.\\

Now we extend the Poisson bracket to more general case. Here we discuss the case of $\{h,\{f,g\}_{P}\}_{P}$ in details, the general case is similar.
As preparation we consider the following Poisson bracket firstly
\begin{equation}
  \{h(w_{\alpha},\zeta_{\beta}),f(u_{\alpha},\xi_{\beta})g(v_{\alpha},\eta_{\beta})\partial_{x}^{\gamma}\partial_{y}^{\sigma}P(x,y)\}_{P},
\end{equation}
where $\gamma,\sigma\in\mathbb{N}^{n}$, and $f(u_{\alpha},\xi_{\beta}),g(v_{\alpha},\eta_{\beta}),h(w_{\alpha},\zeta_{\beta})$ correspond to variables
$x,y,z\in\mathbb{R}^{n}$ respectively. The Poisson brackets (3.5) should be maps as following
$$\mathcal{H}^{\infty}\times(\mathcal{H}^{\infty}\otimes\mathcal{H}^{\infty}\otimes Delta_{x,y})\longrightarrow$$
$$(\mathcal{H}^{\infty}\otimes\mathcal{H}^{\infty}\otimes\mathcal{H}^{\infty}\otimes Delta_{z,x}\otimes Delta_{x,y})\oplus
(\mathcal{H}^{\infty}\otimes\mathcal{H}^{\infty}\otimes\mathcal{H}^{\infty}\otimes Delta_{z,y}\otimes Delta_{x,y}).$$
It is natural that the Poisson bracket (3.5) is defined in the following way
$$\{h(w_{\alpha},\zeta_{\beta}),f(u_{\alpha},\xi_{\beta})g(v_{\alpha},\eta_{\beta})\partial_{x}^{\gamma}\partial_{y}^{\sigma}P(x,y)\}_{P}\doteq$$
\begin{equation}
  [g(v_{\alpha},\eta_{\beta})\{h(w_{\alpha},\zeta_{\beta}),f(u_{\alpha},\xi_{\beta})\}_{P}
\end{equation}
$$+f(u_{\alpha},\xi_{\beta})\{h(w_{\alpha},\zeta_{\beta}),g(v_{\alpha},\eta_{\beta})\}_{P}]\partial_{x}^{\gamma}\partial_{y}^{\sigma}P(x,y)$$
With the help of (3.6) we can define the Poisson bracket $\{h,\{f,g\}_{P}\}_{P}$.
\begin{remark}
From the definition3.2 we know that the operation of Poisson bracket will result in a factor $P(x,y)$.
But, in (3.5) the factor $\partial_{x}^{\gamma}\partial_{y}^{\sigma}P(x,y)$ appears before taking Poisson bracket,
we consider it as a constant in this case. From (3.6) we have
$$\{h(w_{\alpha},\zeta_{\beta}),\partial_{x}^{\gamma}\partial_{y}^{\sigma}P(x,y)\}_{P}=0. $$
\end{remark}

\begin{theorem}
For the Poisson brackets (3.3) and (3.4) the Jacobi identity is valid, which means for the Hamiltonian functions
$f(u_{\alpha},\xi_{\beta}),g(v_{\alpha},\eta_{\beta}),h(w_{\alpha},\zeta_{\beta})$ we have
\begin{equation}
  \{h,\{f,g\}_{P}\}_{P}+\{g,\{h,f\}_{P}\}_{P}+\{f,\{g,h\}_{P}\}_{P}=0.
\end{equation}
Where $f(u_{\alpha},\xi_{\beta}),g(v_{\alpha},\eta_{\beta}),h(w_{\alpha},\zeta_{\beta})$ are assigned to the variables $x,y,z\in\mathbb{R}^{n}$ respectively.
\end{theorem}
\emph{Proof.}
Firstly we discuss the symmetric case that means $P(x,y)$ is symmetric. The left side of (3.7) belongs to
$$(\mathcal{H}^{\infty}\otimes\mathcal{H}^{\infty}\otimes\mathcal{H}^{\infty}\otimes (Delta_{z,x}\oplus Delta_{x,z})\otimes (Delta_{x,y}\oplus Delta_{y,x}))$$
$$\oplus(\mathcal{H}^{\infty}\otimes\mathcal{H}^{\infty}\otimes\mathcal{H}^{\infty}\otimes (Delta_{z,y}\oplus Delta_{y,z})\otimes (Delta_{x,y}\oplus Delta_{y,x})) $$
$$\oplus(\mathcal{H}^{\infty}\otimes\mathcal{H}^{\infty}\otimes\mathcal{H}^{\infty}\otimes (Delta_{z,x}\oplus Delta_{x,z})\otimes (Delta_{z,y}\oplus Delta_{y,z}))$$
We assume that $h(w_{\alpha},\zeta_{\beta}),$ $f(u_{\alpha},\xi_{\beta}),$ $g(v_{\alpha},\eta_{\beta})$ correspond to the first, second and third factors in the tenser $\mathcal{H}^{\infty}\otimes\mathcal{H}^{\infty}\otimes\mathcal{H}^{\infty}$ respectively.
For example, we consider the terms concerning the following tenser space
 $$\mathcal{H}^{\infty}\otimes\mathcal{H}^{\infty}\otimes\mathcal{H}^{\infty}\otimes (Delta_{z,x}\oplus Delta_{x,z})\otimes (Delta_{x,y}\oplus Delta_{y,x}).$$
 The terms of this type are included in $\{h,\{f,g\}_{P}\}_{P}$ and $\{g,\{h,f\}_{P}\}_{P}$. At first we compute $\{h,\{f,g\}_{P}\}_{P}$.
Actually we know that
$$\{h,\{f,g\}_{P}\}_{P}\in(\mathcal{H}^{\infty}\otimes\mathcal{H}^{\infty}\otimes\mathcal{H}^{\infty}\otimes Delta_{z,x}\otimes Delta_{x,y}) $$
 $$\oplus(\mathcal{H}^{\infty}\otimes\mathcal{H}^{\infty}\otimes\mathcal{H}^{\infty}\otimes Delta_{z,y}\otimes Delta_{x,y}). $$
According to (3.6) we have
$$\{h,\{f,g\}_{P}\}_{P}=\{h(w_{\alpha},\zeta_{\beta}),(\partial_{u,x}^{(2,2)}\partial_{\eta,y}^{(3,2)}-\partial_{\xi,x}^{(2,2)}\partial_{v,y}^{(3,2)})
 f(u_{\alpha},\xi_{\beta})g(v_{\alpha},\eta_{\beta})\}_{P}P(x,y)$$
$$=(\partial_{w,z}^{(1,1)}\partial_{\eta,y}^{(3,1)}-\partial_{\zeta,z}^{(1,1)}\partial_{v,y}^{(3,1)})
(\partial_{u,x}^{(2,2)}\partial_{\eta,y}^{(3,2)}-\partial_{\xi,x}^{(2,2)}\partial_{v,y}^{(3,2)})[h(\cdot)f(\cdot)g(\cdot)P(z,y)P(x,y)]$$
$$+(\partial_{w,z}^{(1,1)}\partial_{\xi,x}^{(2,1)}-\partial_{\zeta,z}^{(1,1)}\partial_{u,x}^{(2,1)})
(\partial_{u,x}^{(2,2)}\partial_{\eta,y}^{(3,2)}-\partial_{\xi,x}^{(2,2)}\partial_{v,y}^{(3,2)})[h(\cdot)f(\cdot)g(\cdot)P(z,x)P(x,y)].$$
In previous expression our focus is the term
$$(\partial_{w,z}^{(1,1)}\partial_{\xi,x}^{(2,1)}-\partial_{\zeta,z}^{(1,1)}\partial_{u,x}^{(2,1)})
(\partial_{u,x}^{(2,2)}\partial_{\eta,y}^{(3,2)}-\partial_{\xi,x}^{(2,2)}\partial_{v,y}^{(3,2)})[h(\cdot)f(\cdot)g(\cdot)P(z,x)P(x,y)]. $$
For $\{g,\{h,f\}_{P}\}_{P}$ we have
$$\{g,\{h,f\}_{P}\}_{P}\in(\mathcal{H}^{\infty}\otimes\mathcal{H}^{\infty}\otimes\mathcal{H}^{\infty}\otimes Delta_{z,x}\otimes Delta_{y,x})  $$
$$\oplus(\mathcal{H}^{\infty}\otimes\mathcal{H}^{\infty}\otimes\mathcal{H}^{\infty}\otimes Delta_{z,x}\otimes Delta_{y,z}),  $$
and
$$\{g,\{h,f\}_{P}\}_{P}=\{g(v_{\alpha},\eta_{\beta}),(\partial_{w,z}^{(1,1)}\partial_{\xi,x}^{(2,1)}-\partial_{\zeta,z}^{(1,1)}\partial_{u,x}^{(2,1)})
h(w_{\alpha},\zeta_{\beta})f(u_{\alpha},\xi_{\beta})\}_{P}P(z,x)  $$
$$=(\partial_{v,y}^{(3,2)}\partial_{\zeta,z}^{(1,2)}-\partial_{\eta,y}^{(3,2)}\partial_{w,z}^{(1,2}))
(\partial_{w,z}^{(1,1)}\partial_{\xi,x}^{(2,1)}-\partial_{\zeta,z}^{(1,1)}\partial_{u,x}^{(2,1)})[h(\cdot)f(\cdot)g(\cdot)P(z,x)P(y,z)]  $$
$$+(\partial_{v,y}^{(3,2)}\partial_{\xi,x}^{(2,2)}-\partial_{\eta,y}^{(3,2)}\partial_{u,x}^{(2,2)})
(\partial_{w,z}^{(1,1)}\partial_{\xi,x}^{(2,1)}-\partial_{\zeta,z}^{(1,1)}\partial_{u,x}^{(2,1)})[h(\cdot)f(\cdot)g(\cdot)P(z,x)P(y,x)]  $$
$$=(\partial_{v,y}^{(3,2)}\partial_{\zeta,z}^{(1,2)}-\partial_{\eta,y}^{(3,2)}\partial_{w,z}^{(1,2}))
(\partial_{w,z}^{(1,1)}\partial_{\xi,x}^{(2,1)}-\partial_{\zeta,z}^{(1,1)}\partial_{u,x}^{(2,1)})[h(\cdot)f(\cdot)g(\cdot)P(z,x)P(y,z)]  $$
$$-(\partial_{w,z}^{(1,1)}\partial_{\xi,x}^{(2,1)}-\partial_{\zeta,z}^{(1,1)}\partial_{u,x}^{(2,1)})
(\partial_{u,x}^{(2,2)}\partial_{\eta,y}^{(3,2)}-\partial_{\xi,x}^{(2,2)}\partial_{v,y}^{(3,2)})[h(\cdot)f(\cdot)g(\cdot)P(z,x)P(x,y)].  $$
Thus the terms concerning the factor $P(z,x)P(x,y)$ on the left side of (3.7) should be canceled out each other. By the similar computation we know the previous discussion is valid for the terms of other types.\\
For the anti-symmetric case we discuss the terms concerning $P(z,x)P(x,y)$ in details also. We have
$$\{h,\{f,g\}_{P}\}_{P}=\{h(w_{\alpha},\zeta_{\beta}),(\partial_{u,x}^{(2,2)}\partial_{\eta,y}^{(3,2)}+\partial_{\xi,x}^{(2,2)}\partial_{v,y}^{(3,2)})
 f(u_{\alpha},\xi_{\beta})g(v_{\alpha},\eta_{\beta})\}_{P}P(x,y)$$
$$=(\partial_{w,z}^{(1,1)}\partial_{\eta,y}^{(3,1)}+\partial_{\zeta,z}^{(1,1)}\partial_{v,y}^{(3,1)})
(\partial_{u,x}^{(2,2)}\partial_{\eta,y}^{(3,2)}+\partial_{\xi,x}^{(2,2)}\partial_{v,y}^{(3,2)})[h(\cdot)f(\cdot)g(\cdot)P(z,y)P(x,y)]$$
$$+(\partial_{w,z}^{(1,1)}\partial_{\xi,x}^{(2,1)}+\partial_{\zeta,z}^{(1,1)}\partial_{u,x}^{(2,1)})
(\partial_{u,x}^{(2,2)}\partial_{\eta,y}^{(3,2)}+\partial_{\xi,x}^{(2,2)}\partial_{v,y}^{(3,2)})[h(\cdot)f(\cdot)g(\cdot)P(z,x)P(x,y)].$$
$$\{g,\{h,f\}_{P}\}_{P}=\{g(v_{\alpha},\eta_{\beta}),(\partial_{w,z}^{(1,1)}\partial_{\xi,x}^{(2,1)}+\partial_{\zeta,z}^{(1,1)}\partial_{u,x}^{(2,1)})
h(w_{\alpha},\zeta_{\beta})f(u_{\alpha},\xi_{\beta})\}_{P}P(z,x)  $$
$$=(\partial_{v,y}^{(3,2)}\partial_{\zeta,z}^{(1,2)}+\partial_{\eta,y}^{(3,2)}\partial_{w,z}^{(1,2}))
(\partial_{w,z}^{(1,1)}\partial_{\xi,x}^{(2,1)}+\partial_{\zeta,z}^{(1,1)}\partial_{u,x}^{(2,1)})[h(\cdot)f(\cdot)g(\cdot)P(z,x)P(y,z)]  $$
$$-(\partial_{w,z}^{(1,1)}\partial_{\xi,x}^{(2,1)}+\partial_{\zeta,z}^{(1,1)}\partial_{u,x}^{(2,1)})
(\partial_{u,x}^{(2,2)}\partial_{\eta,y}^{(3,2)}+\partial_{\xi,x}^{(2,2)}\partial_{v,y}^{(3,2)})[h(\cdot)f(\cdot)g(\cdot)P(z,x)P(x,y)].  $$
Thus we have same conclusion.
$\blacksquare$

\subsection{The star products on $\mathcal{H}^{\infty}$}
Same as the finite dimensional case we consider the set of formal power series $\sum_{k\geq0}\hbar^{k}f_{k}(u_{\alpha},\xi_{\beta})$
denoted by
\begin{equation}
  \mathcal{H}_{\hbar}^{\infty}=\{\sum_{k\geq0}\hbar^{k}f_{k}(u_{\alpha},\xi_{\beta})|
  f_{k}(u_{\alpha},\xi_{\beta})\in\mathcal{H}^{\infty},k=0,1,\cdots \}.
\end{equation}
Where $\hbar$ is a parameter playing the role of Planck constant usually. Under the usual pointwise multiplication of power series $\mathcal{H}_{\hbar}^{\infty}$ is an algebra.

\begin{definition}
  Let $P(x,y)\in Delta_{x,y}$, $f(u_{\alpha},\xi_{\beta}),g(v_{\alpha},\eta_{\beta})\in\mathcal{H}^{\infty}$, where
  $f(u_{\alpha},\xi_{\beta})$, $g(v_{\alpha},\eta_{\beta})$ are assigned to variables $x,y\in\mathbb{R}^{n}$, we define
\begin{equation}
\begin{array}{c}
   f(u_{\alpha},\xi_{\beta})\star_{P} g(v_{\alpha},\eta_{\beta})\doteq f(u_{\alpha},\xi_{\beta})g(v_{\alpha},\eta_{\beta})+ \\
   \sum_{k\geq1}\frac{\hbar^{k}}{k!}(\partial_{u,x}\partial_{\eta,y}\mp\partial_{\xi,x}\partial_{v,y})^{k}
   (f(u_{\alpha},\xi_{\beta})g(v_{\alpha},\eta_{\beta})P(x,y)).
\end{array}
\end{equation}
For two formal power series $\sum_{k\geq0}\hbar^{k}f_{k}(u_{\alpha},\xi_{\beta}),\sum_{k\geq0}\hbar^{k}g_{k}(v_{\alpha},\eta_{\beta})\in\mathcal{H}_{\hbar}^{\infty}$,
their star product is defined to be
\begin{equation}
\begin{array}{c}
  (\sum_{k\geq0}\hbar^{k}f_{k}(u_{\alpha},\xi_{\beta}))\star_{P}(\sum_{l\geq0}\hbar^{l}g_{l}(v_{\alpha},\eta_{\beta})) \\
  \doteq\sum_{k,l\geq0}\hbar^{k+l}f_{k}(u_{\alpha},\xi_{\beta})\star_{P}g_{l}(v_{\alpha},\eta_{\beta})
\end{array}
\end{equation}
\end{definition}

To simplify the expression in (3.9) we introduce some formal notations as following. Let
$\sigma_{u,v;x,y}^{\mp}$ be an operator such that
$$[\sigma_{u,v;x,y}^{\mp}]^{k}(f(u_{\alpha},\xi_{\beta})g(v_{\alpha},\eta_{\beta}))$$
$$\doteq(\partial_{u,x}\partial_{\eta,y}\mp\partial_{\xi,x}\partial_{v,y})^{k}(f(u_{\alpha},\xi_{\beta})g(v_{\alpha},\eta_{\beta})P(x,y))
\quad (k\geq1),$$
and
$$[\sigma_{u,v;x,y}^{\mp}]^{0}(f(u_{\alpha},\xi_{\beta})g(v_{\alpha},\eta_{\beta}))
=f(u_{\alpha},\xi_{\beta})g(v_{\alpha},\eta_{\beta}),$$
then (3.9) can be rewritten to be
\begin{equation}
  f(u_{\alpha},\xi_{\beta})\star_{P} g(v_{\alpha},\eta_{\beta})=\exp(\hbar{\sigma_{u,v;x,y}^{\mp}})
[f(u_{\alpha},\xi_{\beta})g(v_{\alpha},\eta_{\beta})],
\end{equation}
where
$$\exp(\hbar{\sigma_{u,v;x,y}^{\mp}})\doteq \sum_{k\geq0 }\frac{\hbar^{k}}{k!}[\sigma_{u,v;x,y}^{\mp}]^{k}.$$

\begin{remark}
The star products (3.9), (3.10) is not usual algebraic operation, which are maps
$$\mathcal{H}_{\hbar}^{\infty}\times\mathcal{H}_{\hbar}^{\infty}\rightarrow\mathcal{H}_{\hbar}^{\infty}\otimes\mathcal{H}_{\hbar}^{\infty}\otimes Delta_{x,y}.$$
\end{remark}
\begin{remark}
It is obvious that the star products (3.9) are non-commutative, and in the case of minus sign we have
$$f(u_{\alpha},\xi_{\beta})\star_{P}g(v_{\alpha},\eta_{\beta})-g(v_{\alpha},\eta_{\beta})\star_{P}f(u_{\alpha},\xi_{\beta})$$
$$=\hbar[(\partial_{u,x}\partial_{\eta,y}-\partial_{\xi,x}\partial_{v,y})f(u_{\alpha},\xi_{\beta})g(v_{\alpha},\eta_{\beta})(P(x,y)+P(y,x))]  $$
$$+\; \textit{high oeder terms of}\;\hbar.$$
Noting that $P(x,y)+P(y,x)$ is symmetric, thus the semi-classical limit of the star products (3.9) is Poisson bracket defined in definition3.2 in the case of minus sign. The case of plus sign is similar.
\end{remark}

More generally, we need to extend the star product to the case of three or more Hamiltonian functions. For example, we discuss the case of
three Hamiltonian functions in details, let $f(u_{\alpha},\xi_{\beta}),g(v_{\alpha},\eta_{\beta}),h(w_{\alpha},\zeta_{\beta})\in\mathcal{H}^{\infty}$,
they are assigned to variables $x,y,z\in\mathbb{R}^{n}$ respectively. We want to discuss the following star products
\begin{equation}
  h(w_{\alpha},\zeta_{\beta})\star_{P}(f(u_{\alpha},\xi_{\beta})g(v_{\alpha},\eta_{\beta})),
\end{equation}
\begin{equation}
  (h(w_{\alpha},\zeta_{\beta})f(u_{\alpha},\xi_{\beta}))\star_{P}g(v_{\alpha},\eta_{\beta}),
\end{equation}
\begin{equation}
  h(w_{\alpha},\zeta_{\beta})\star_{P}f(u_{\alpha},\xi_{\beta})\star_{P}g(v_{\alpha},\eta_{\beta}).
\end{equation}
 The star products (3.12),(3.13) and (3.14) should concern the maps at different levels as followings
$$\mathcal{H}_{\hbar}^{\infty}\times(\mathcal{H}_{\hbar}^{\infty}\otimes\mathcal{H}_{\hbar}^{\infty})\rightarrow
 \mathcal{H}_{\hbar}^{\infty}\otimes\mathcal{H}_{\hbar}^{\infty}\otimes\mathcal{H}_{\hbar}^{\infty}\otimes Delte_{z,x}\otimes Delta_{z,y}, $$
$$(\mathcal{H}_{\hbar}^{\infty}\otimes\mathcal{H}_{\hbar}^{\infty})\times\mathcal{H}_{\hbar}^{\infty}\rightarrow
 \mathcal{H}_{\hbar}^{\infty}\otimes\mathcal{H}_{\hbar}^{\infty}\otimes\mathcal{H}_{\hbar}^{\infty}\otimes Delte_{z,y}\otimes Delta_{x,y}, $$
$$\mathcal{H}_{\hbar}^{\infty}\times\mathcal{H}_{\hbar}^{\infty}\times\mathcal{H}_{\hbar}^{\infty}\rightarrow
\mathcal{H}_{\hbar}^{\infty}\otimes\mathcal{H}_{\hbar}^{\infty}\otimes\mathcal{H}_{\hbar}^{\infty}\otimes Delte_{z,x}\otimes Delta_{z,y}\otimes
Delta_{x,y}.$$

Because the star products are non-commutative we need to make setting for order of factors in the tenser space, more precisely, as an example we describe that in the case of $\mathcal{H}_{\hbar}^{\infty}\times\mathcal{H}_{\hbar}^{\infty}\times\mathcal{H}_{\hbar}^{\infty}$ as following
$$\overset{h(w_{\alpha},\zeta_{\beta})}{\overbrace{\mathcal{H}_{\hbar}^{\infty}}}\times
\overset{f(u_{\alpha},\xi_{\beta})}{\overbrace{\mathcal{H}_{\hbar}^{\infty}}}\times
\overset{g(v_{\alpha},\eta_{\beta})}{\overbrace{\mathcal{H}_{\hbar}^{\infty}}}.  $$
The cases of $\mathcal{H}_{\hbar}^{\infty}\times(\mathcal{H}_{\hbar}^{\infty}\otimes\mathcal{H}_{\hbar}^{\infty})$
and $(\mathcal{H}_{\hbar}^{\infty}\otimes\mathcal{H}_{\hbar}^{\infty})\times\mathcal{H}_{\hbar}^{\infty}$ are same.

Now we define the star product (3.12) to be
\begin{subequations}
\begin{equation}
\begin{array}{c}
  h(w_{\alpha},\zeta_{\beta})\star_{P}(f(u_{\alpha},\xi_{\beta})g(v_{\alpha},\eta_{\beta})) \\
  \doteq \exp(\hbar\sigma_{w,v;z,y}^{\mp})[\exp(\hbar\sigma_{w,u;z,x}^{\mp})(h(\cdot)f(\cdot)g(\cdot))],
\end{array}
\end{equation}
or,
\begin{equation}
\begin{array}{c}
  h(w_{\alpha},\zeta_{\beta})\star_{P}(f(u_{\alpha},\xi_{\beta})g(v_{\alpha},\eta_{\beta})) \\
  \doteq \exp(\hbar\sigma_{w,u;z,x}^{\mp})[\exp(\hbar\sigma_{w,v;z,y}^{\mp})(h(\cdot)f(\cdot)g(\cdot))],
\end{array}
\end{equation}
\end{subequations}

Similarly, the star product (3.13) is defined to be
\begin{subequations}
\begin{equation}
  \begin{array}{c}
    (h(w_{\alpha},\zeta_{\beta})f(u_{\alpha},\xi_{\beta}))\star_{P}g(v_{\alpha},\eta_{\beta}) \\
    \doteq\exp(\hbar\sigma_{w,v;z,y}^{\mp})[\exp(\hbar\sigma_{u,v;x,y}^{\mp})(h(\cdot)f(\cdot)g(\cdot))],
  \end{array}
\end{equation}
or,
\begin{equation}
\begin{array}{c}
  (h(w_{\alpha},\zeta_{\beta})f(u_{\alpha},\xi_{\beta}))\star_{P}g(v_{\alpha},\eta_{\beta}) \\
  \doteq \exp(\hbar\sigma_{u,v;x,y}^{\mp})[\exp(\hbar\sigma_{w,v;z,y}^{\mp})(h(\cdot)f(\cdot)g(\cdot))].
\end{array}
\end{equation}
\end{subequations}

The following propsition shows that the definitions of the star products (3.12), (3.13) mentioned above are well defined.
\begin{propsition}
The following formulas are valid
\begin{equation}
\begin{array}{c}
  \exp(\hbar\sigma_{w,v;z,y}^{\mp})[\exp(\hbar\sigma_{w,u;z,x}^{\mp})(h(\cdot)f(\cdot)g(\cdot))] \\
  =\exp(\hbar\sigma_{w,u;z,x}^{\mp})[\exp(\hbar\sigma_{w,v;z,y}^{\mp})(h(\cdot)f(\cdot)g(\cdot))]
\end{array}
\end{equation}
\begin{equation}
\begin{array}{c}
  \exp(\hbar\sigma_{w,v;z,y}^{\mp})[\exp(\hbar\sigma_{u,v;x,y}^{\mp})(h(\cdot)f(\cdot)g(\cdot))] \\
  =\exp(\hbar\sigma_{u,v;x,y}^{\mp})[\exp(\hbar\sigma_{w,v;z,y}^{\mp})(h(\cdot)f(\cdot)g(\cdot))]
\end{array}
\end{equation}
Here $f(u_{\alpha},\xi_{\beta}),g(v_{\alpha},\eta_{\beta}),h(w_{\alpha},\zeta_{\beta})\in\mathcal{H}^{\infty}$ are same as mentioned above.
\end{propsition}

\emph{Proof.}$\;$ Here we check only the formula (3.17). The proof of (3.18) is similar. For the formula (3.17)
it is enough for us to prove the following equality
$$\exp(\hbar\sigma_{w,v;z,y}^{\mp})\cdot\exp(\hbar\sigma_{w,u;z,x}^{\mp})
=\exp(\hbar\sigma_{w,u;z,x}^{\mp})\cdot\exp(\hbar\sigma_{w,v;z,y}^{\mp}),$$
or, equivalently,
$$[\sigma_{w,v;z,y}^{\mp}]^{k}[\sigma_{w,u;z,x}^{\mp}]^{l}=[\sigma_{w,u;z,x}^{\mp}]^{l}[\sigma_{w,v;z,y}^{\mp}]^{k}.$$

Actually, the operators $[\sigma_{w,v;z,y}^{\mp}]^{k}$ and $[\sigma_{w,u;z,x}^{\mp}]^{l}$
act on the tenser space
$$\mathcal{H}^{\infty}\otimes\mathcal{H}^{\infty}\otimes\mathcal{H}^{\infty}\otimes Delta_{z,x}\otimes Delta_{z,y}.$$
For convenience we make the order of factors in above tenser space, which can be shown with variables in the following diagram
$$(w,\zeta)\rightarrow(u,\xi)\rightarrow(v,\eta)\rightarrow(z,x)\rightarrow(z,y).$$
Thus the operator $[\sigma_{w,u;z,x}^{\mp}]^{l}$ concerns the first, second and fourth factors in above tenser space respectively,
and the operator $[\sigma_{w,v;z,y}^{\mp}]^{k}$ concerns the first, third and fifth factors respectively.

Up to now, it is sufficient for us to check the following equality
$$(\partial_{w,z}^{(1,4)}\partial_{\xi,x}^{(2,4)}\mp\partial_{\zeta,z}^{(1,4)}\partial_{u,x}^{(2,4)})^{k}
(\partial_{w,z}^{(1,5)}\partial_{\eta,y}^{(3,5)}\mp\partial_{\zeta,z}^{(1,5)}\partial_{v,y}^{(3,5)})^{l}[hfgP(z,x)P(z,y)]$$
$$=(\partial_{w,z}^{(1,5)}\partial_{\eta,y}^{(3,5)}\mp\partial_{\zeta,z}^{(1,5)}\partial_{v,y}^{(3,5)})^{l}
(\partial_{w,z}^{(1,4)}\partial_{\xi,x}^{(2,4)}\mp\partial_{\zeta,z}^{(1,4)}\partial_{u,x}^{(2,4)})^{k}[hfgP(z,x)P(z,y)].$$
where the indices $(i,j)$ in above expressions correspond to the positions of factors in the tenser space. Above equality is valid obviously.   $\blacksquare$

Base on the above discussion, we can define the star product in (3.14) in the following way
\begin{subequations}
\begin{equation}
\begin{array}{c}
  h(w_{\alpha},\zeta_{\beta})\star_{P}f(u_{\alpha},\xi_{\beta})\star_{P}g(v_{\alpha},\eta_{\beta}) \\
  \doteq h(w_{\alpha},\zeta_{\beta})\star_{P}(f(u_{\alpha},\xi_{\beta})\star_{P}g(v_{\alpha},\eta_{\beta})),
\end{array}
\end{equation}
or
\begin{equation}
\begin{array}{c}
  h(w_{\alpha},\zeta_{\beta})\star_{P}f(u_{\alpha},\xi_{\beta})\star_{P}g(v_{\alpha},\eta_{\beta}) \\
  \doteq(h(w_{\alpha},\zeta_{\beta})\star_{P}f(u_{\alpha},\xi_{\beta}))\star_{P}g(v_{\alpha},\eta_{\beta}).
\end{array}
\end{equation}
\end{subequations}

\begin{theorem}
We have
\begin{equation}
\begin{array}{c}
  h(w_{\alpha},\zeta_{\beta})\star_{P}(f(u_{\alpha},\xi_{\beta})\star_{P}g(v_{\alpha},\eta_{\beta})) \\
  =(h(w_{\alpha},\zeta_{\beta})\star_{P}f(u_{\alpha},\xi_{\beta}))\star_{P}g(v_{\alpha},\eta_{\beta}).
\end{array}
\end{equation}
\end{theorem}
\emph{Proof.}$\;$ To prove formula (3.20) we need to prove
$$\exp(\hbar\sigma_{w,v;z,y}^{\mp})\exp(\hbar\sigma_{w,u;z,x}^{\mp})
[h(\cdot)\exp(\hbar\sigma_{u,v;x,y}^{\mp})(f(\cdot)g(\cdot))]$$
$$=\exp(\hbar\sigma_{w,v;z,y}^{\mp})\exp(\hbar\sigma_{u,v;x,y}^{\mp})
[(\exp(\hbar\sigma_{w,u;z,x}^{\mp})h(\cdot)f(\cdot))g(\cdot)].$$

By a straightforward calculation as what we do in the proof of propsition3.1 we can check the following equality
$$\exp(\hbar\sigma_{w,u;z,x}^{\mp})[h(\cdot)\exp(\hbar\sigma_{u,v;x,y}^{\mp})(f(\cdot)g(\cdot))]$$
$$=\exp(\hbar\sigma_{u,v;x,y}^{\mp})[(\exp(\hbar\sigma_{w,u;z,x}^{\mp})h(\cdot)f(\cdot))g(\cdot)]$$
is valid. $\blacksquare$

The formula (3.20) means that the star products defined above satisfy the associative law.

Furthermore, we generalize the star product to the situation with more factors. Let
$$f_{1}(u_{\alpha}^{(1)},\xi_{\beta}^{(1)}),\cdots,f_{p}(u_{\alpha}^{(p)},\xi_{\beta}^{(p)}),
g_{1}(v_{\alpha}^{(1)},\eta_{\beta}^{(1)}),\cdots,g_{q}(v_{\alpha}^{(q)},\eta_{\beta}^{(q)}),$$
($p+q=m$) be $m$ Hamiltonian functions, where we assign $f_{i}(u_{\alpha}^{(i)},\xi_{\beta}^{(i)})$ to variables $x_{i}\in\mathbb{R}^{n}$, $g_{j}(v_{\alpha}^{(j)},\eta_{\beta}^{(j)})$
to variables $y_{j}\in\mathbb{R}^{n}$ $(i=1,\cdots,p;j=1,\cdots,q)$.
Now we define the following star product in the way similar to (3.15) and (3.16)
\begin{equation}
  \begin{array}{c}
    [f_{1}(u_{\alpha}^{(1)},\xi_{\beta}^{(1)})\cdots f_{p}(u_{\alpha}^{(p)},\xi_{\beta}^{(p)})]\star
    [g_{1}(v_{\alpha}^{(1)},\eta_{\beta}^{(1)})\cdots g_{q}(v_{\alpha}^{(q)},\eta_{\beta}^{(q)})] \\
    \doteq\prod_{1\leq i\leq p,1\leq j\leq q}\exp(\hbar\sigma_{u^{(i)},v^{(j)};x_{i},y_{j}}^{\mp})
    (f_{1}\cdots f_{p}g_{1}\cdots g_{q}).
  \end{array}
\end{equation}

Similar to (3.19) we have
\begin{equation}
  \begin{array}{c}
    f_{1}(u_{\alpha}^{(1)},\xi_{\beta}^{(1)})\star\cdots \star f_{p}(u_{\alpha}^{(p)},\xi_{\beta}^{(p)}) \\
    \doteq\prod_{1\leq i<j\leq p}\exp(\hbar\sigma_{u^{(i)},u^{(j)};x_{i},x_{j}}^{\mp})(f_{1}\cdots f_{p}).
  \end{array}
\end{equation}

\section{The star products on $\mathcal{H}_{den}$ and $H^{\infty}$}

\subsection{The Poisson brackets on $\mathcal{H}_{den}$ and $H^{\infty}$}

Taking distributions $P(x,y)\in Delta_{x,y}$ as in definition3.1 and definition3.2, we have following definitions,
\begin{definition}
For two Hamilton densities $f(\partial_{x}^{\alpha}\varphi(x),\partial_{x}^{\beta}\pi(x)),g(\partial_{y}^{\alpha}\varphi(y),\partial_{y}^{\beta}\pi(y))$,
their Poisson bracket is defined to be
\begin{equation}
\begin{array}{c}
  \{f(\partial_{x}^{\alpha}\varphi(x),\partial_{x}^{\beta}\pi(x)),g(\partial_{y}^{\alpha}\varphi(y),\partial_{y}^{\beta}\pi(y))\}_{P} \\
  =\{f(u_{\alpha},\xi_{\beta}),g(v_{\alpha},\eta_{\beta})\}_{P}|_{u_{\alpha}=\partial_{x}^{\alpha}\varphi(x),
 \xi_{\beta}=\partial_{x}^{\beta}\pi(x),v_{\alpha}=\partial_{y}^{\alpha}\varphi(y),\eta_{\beta}=\partial_{y}^{\beta}\pi(y)}.
\end{array}
\end{equation}
\end{definition}
\begin{definition}
For $F(\varphi,\pi)\in H^{\infty}$, $g(\partial_{y}^{\alpha}\varphi(y),\partial_{y}^{\beta}\pi(y))\in\mathcal{H}_{den}$,
their Poisson bracket is defined to be
\begin{equation}
  \{F(\varphi,\pi),g(\partial_{y}^{\alpha}\varphi(y),\partial_{y}^{\beta}\pi(y))\}_{P}=\langle\{f,g\}_{P},1_{x}\rangle.
\end{equation}
Where
$$F(\varphi,\pi)=\int_{\mathbb{R}^{n}}f(\partial_{x}^{\alpha}\varphi(x),\partial_{x}^{\beta}\pi(x))dx,$$
and $f(u_{\alpha},\xi_{\beta})\in\mathcal{H}^{\infty}$ satisfies the condition $\mathbf{B}$, $\{f,g\}_{P}$ is defined in definition4.1.
\end{definition}
\begin{definition}
For two functionals in $H^{\infty}$
$$F(\varphi,\pi)=\int_{\mathbb{R}^{n}}f(\partial_{x}^{\alpha}\varphi(x),\partial_{x}^{\beta}\pi(x))dx,$$
$$G(\varphi,\pi)=\int_{\mathbb{R}^{n}}g(\partial_{y}^{\alpha}\varphi(y),\partial_{y}^{\beta}\pi(y))dy, $$
where $f(u_{\alpha},\xi_{\beta}),g(\partial_{y}^{\alpha}\varphi(y),\partial_{y}^{\beta}\pi(y))\in\mathcal{H}^{\infty}$ satisfy the condition $\mathbf{B}$, their Poisson bracket is defined to be
\begin{equation}
  \{F(\varphi,\pi),G(\varphi,\pi)\}_{P}=\int_{\mathbb{R}^{n}}\langle\{f,g\}_{P},1_{x}\rangle dy.
\end{equation}
\end{definition}

It is obvious that we have
$$\{F,G\}_{P}=\int_{\mathbb{R}^{n}}\{F,g\}_{P}dy.$$

\begin{remark}
The Poisson brackets defined as above are the following maps
$$\mathcal{H}_{den}\times\mathcal{H}_{den}\rightarrow\mathcal{H}_{den}\otimes\mathcal{H}_{den}\otimes Delta_{x,y}, $$
$$H^{\infty}\times\mathcal{H}_{den}\rightarrow\mathcal{H}_{den},$$
$$H^{\infty}\times H^{\infty}\rightarrow H^{\infty}.$$
\end{remark}
\begin{remark}
From definition3.1 we have
$$\{\varphi(x),\pi(y)\}_{P}=P(x,y),\{\varphi(x),\varphi(y)\}_{P}=0,\{\pi(x),\pi(y)\}_{P}=0,$$
where $P(x,y)$ is symmetric or anti-symmetric. We call above Poisson brackets the basic Poisson brackets.

Recalling C.Gardner \cite{b}, V.E.Zakharov, L.D.Faddeev \cite{c}, and F.Magri \cite{d}, we know that in the case of KdV equation
there are two types of Poisson brackets, they are
$\{u(x),u(y)\}_{1}=\partial_{y}\delta(x-y)$(\cite{b},\cite{c}), and
$\{u(x),u(y)\}_{2}=(u(x)+u(y))\partial_{y}\delta(x-y)-\frac{\alpha^{2}}{2}\partial_{y}^{3}\delta(x-y)$ (\cite{d}). Though the Poisson
brackets of KdV equation are different from ones in this paper, but the anti-symmetric distributions appear indeed.
\end{remark}
\begin{remark}
It is easy to check that Poisson brackets (4.1), (4.2) and (4.3) are anti-symmetric, bilinear and are derivatives for first and second variables respectively. But $H^{\infty}$ is a Poisson algebra really.
\end{remark}
\begin{prop}
For Poisson bracket (4.2) we have
\begin{equation}
  \langle\{f,g\}_{P},1_{x}\rangle=\partial_{\eta,y}(g,y)\langle P(x,y),\frac{\delta F}{\delta\varphi}\rangle
  \mp\partial_{v,y}(g,y)\langle P(x,y),\frac{\delta F}{\delta\pi}\rangle.
\end{equation}
\end{prop}
\emph{Proof.}$\;$By definition of Poisson bracket (4.2) we have
$$\langle\{f,g\}_{P},1_{x}\rangle=\langle[\partial_{u,x}(f,x)\partial_{\eta,y}(g,y)\mp
 \partial_{\xi,x}(f,x)\partial_{v,y}(g,y)]P(x,y),1_{x}\rangle $$
$$=\partial_{\eta,y}(g,y)\langle\partial_{u,x}(f,x)P(x,y),1_{x} \rangle\mp
\partial_{v,y}(g,y)\langle\partial_{\xi,x}(f,x)P(x,y),1_{x} \rangle.  $$
and
$$\langle\partial_{u,x}(f,x)P(x,y),1_{x}\rangle=\langle P(x,y),\partial_{u,x}^{t}(f,x)(1_{x})\rangle  $$
$$=\langle P(x,y),L_{u,x}f(\partial_{x}^{\alpha}\varphi(x),\partial_{x}^{\beta}\pi(x))\rangle,$$
$$\langle\partial_{\xi,x}(f,x)P(x,y),1_{x} \rangle=\langle P(x,y),L_{\xi,x}f(\partial_{x}^{\alpha}\varphi(x),\partial_{x}^{\beta}\pi(x))\rangle.$$
Recalling the formulas (2.17),(2.18) we get the formula (4.4).$\blacksquare$
\begin{remark}
Particularly, in the case of $P(x,y)=\delta(x-y)$, the formula (4.4) can be expressed as following
\begin{equation}
 \{F(\varphi,\pi),g(\partial_{y}^{\alpha}\varphi(y),\partial_{y}^{\beta}\pi(y))\}_{\delta}
 =\partial_{\eta,y}(g,y)\frac{\delta F}{\delta\varphi}-\partial_{v,y}(g,y)\frac{\delta F}{\delta\pi}.
\end{equation}
\end{remark}
\begin{prop}
For Poisson bracket (4.3) we have
\begin{equation}
  \{F(\varphi,\pi),G(\varphi,\pi)\}_{P}=\langle P(x,y),\frac{\delta F}{\delta\varphi}\frac{\delta G}{\delta\pi}
  \mp\frac{\delta F}{\delta\pi}\frac{\delta G}{\delta\varphi}\rangle.
\end{equation}
\end{prop}
\emph{Proof.}$\;$ Because $f(u_{\alpha},\xi_{\beta}),g(v_{\alpha},\eta_{\beta})$ satisfy the condition $\mathbf{B}$,
in a neighborhood of origin we have
$$|f(u_{\alpha},\xi_{\beta})|\leq C\sum(|u_{\alpha}|+|\xi_{\beta}|),
 |g(v_{\alpha},\eta_{\beta})|\leq C\sum(|v_{\alpha}|+|\eta_{\beta}|). $$
Thus£¬it is easy to check that
$$\partial_{u,x}^{t}(f,x)(1_{x}),\partial_{\xi,x}^{t}(f,x)(1_{x}),\partial_{v,x}^{t}(g,x)(1_{x}),\partial_{\eta,x}^{t}(g,x)(1_{x})
\in \mathcal{S}(\mathbb{R}^{n}). $$
According to (4.4) we have
$$\{F(\varphi,\pi),G(\varphi,\pi)\}_{P}$$
$$=\int_{\mathbb{R}^{n}}(\partial_{\eta,y}(g,y)\langle P(x,y),\frac{\delta F}{\delta\varphi}\rangle
  \mp\partial_{v,y}(g,y)\langle P(x,y),\frac{\delta F}{\delta\pi}\rangle)dy. $$
By part of integral we have
$$\int_{\mathbb{R}^{n}}\partial_{\eta,y}(g,y)\langle P(x,y),\frac{\delta F}{\delta\varphi}\rangle dy
=\int_{\mathbb{R}^{n}}\partial_{\eta,y}^{t}(g,y)(1_{y})\langle P(x,y),\frac{\delta F}{\delta\varphi}\rangle dy  $$
$$=\int_{\mathbb{R}^{n}}\frac{\delta G}{\delta\pi}\langle P(x,y),\frac{\delta F}{\delta\varphi}dy \rangle,$$
and
$$\int_{\mathbb{R}^{n}}\partial_{v,y}(g,y)\langle P(x,y),\frac{\delta F}{\delta\pi}\rangle dy
=\int_{\mathbb{R}^{n}}\frac{\delta G}{\delta\varphi}\langle P(x,y),\frac{\delta F}{\delta\pi}\rangle dy. $$
Thus
$$\{F(\varphi,\pi),G(\varphi,\pi)\}_{P}=\langle P(x,y),\frac{\delta F}{\delta\varphi}\frac{\delta G}{\delta\pi}
  \mp\frac{\delta F}{\delta\pi}\frac{\delta G}{\delta\varphi}\rangle. $$\\
$\blacksquare$
\begin{remark}
From proposition4.2 we have
\begin{equation}
  \int_{\mathbb{R}^{n}}\langle\{f,g\}_{P},1_{x}\rangle dy
  =\int_{\mathbb{R}^{n}}\langle\{f,g\}_{P},1_{y}\rangle dx.
\end{equation}
Thus the Poisson bracket (4.3) is well defined. In the case of $P(x,y)=\delta(x-y)$ we have
\begin{equation}
  \{F(\varphi,\pi),G(\varphi,\pi)\}_{\delta}=\int_{\mathbb{R}^{n}}(\frac{\delta F}{\delta\varphi}\frac{\delta G}{\delta\pi}
  -\frac{\delta F}{\delta\pi}\frac{\delta G}{\delta\varphi})dx.
\end{equation}
It is obvious that the expressions in (4.1), (4.2) and (4.3) satisfy all of conditions which are needed for Poisson bracket. The Poisson brackets (3.3), (3.4) and (4.1) , (4.2), (4.3)are compatible.
\end{remark}

In order to discuss the star products between $\mathcal{H}_{den}$ and $H^{\infty}$, we need to consider the structure of module of
$H^{\infty}\otimes\mathcal{H}_{den}$. Actually, we can define a multiplication on $H^{\infty}\otimes\mathcal{H}_{den}$ in a natural way.
let $g(\partial_{x}^{\alpha}\varphi(x),\partial_{x}^{\beta}\pi(x))\in\mathcal{H}_{den}$, $F(\varphi,\pi)\in H^{\infty}$,
we define the multiplication as following
\begin{equation}
  (F(\varphi,\pi),g(\partial_{x}^{\alpha}\varphi(x),\partial_{x}^{\beta}\pi(x)))\mapsto
F(\varphi,\pi)g(\partial_{x}^{\alpha}\varphi(x),\partial_{x}^{\beta}\pi(x)).
\end{equation}
Under the multiplication (4.9) $H^{\infty}\otimes\mathcal{H}_{den}$ becomes a $\mathcal{H}_{den}$ module. Similarly,
$H^{\infty}\otimes\mathcal{H}_{den}\otimes Delta_{x,y}$ is a  $\mathcal{H}_{den}$ module also. Now we have
\begin{definition}
For Hamiltonian densities $g(\partial_{y}^{\alpha}\varphi(y),\partial_{y}^{\beta}\pi(y)))$,
$h(\partial_{z}^{\alpha}\varphi(z),\partial_{z}^{\beta}\pi(z)))$, and Hamiltonian functionals $F(\varphi,\pi),H(\varphi,\pi)$,
we define
\begin{equation}
  \{h,gF\}_{P}\doteq g\{h,F\}_{P}+F\{h,g\}_{P},
\end{equation}
and
\begin{equation}
  \{H,gF\}_{P}\doteq g\{H,F\}_{P}+F\{H,g\}_{P}.
\end{equation}
\end{definition}

\begin{remark}
We can discuss the multiple Poisson brackets in the cases of fields and functionals also. For example, we have
$$\{\{h,F\}_{P},G\}_{P}\doteq\langle\langle \{\{h,f\}_{P},g\}_{P},1_{x}\rangle,1_{y}\rangle,$$
it is easy to check that above definition is well defined. In the same way as one in section3 we can discuss the multiple Poisson brackets on $H^{\infty}\otimes\mathcal{H}_{den}$.
\end{remark}

\subsection{The star products on $\mathcal{H}_{den}$ and $H^{\infty}$}

Same as what we did in section3 we consider the algebras consisting of the formal power series similar to ones in section3 with coefficients in
$\mathcal{H}_{den}$ or $H^{\infty}$. They are denoted as followings
$$\mathcal{H}_{den,\hbar}=\{\sum_{k\geq0}\hbar^{k}f_{k}(\partial_{x}^{\alpha}\varphi,\partial_{x}^{\beta}\pi)
|f_{k}(\partial_{x}^{\alpha}\varphi,\partial_{x}^{\beta}\pi)\in\mathcal{H}_{den},k=0,1,\cdots \}, $$
$$H_{\hbar}^{\infty}=\{\sum_{k\geq0}\hbar^{k}F_{k}(\varphi,\pi)|F_{k}(\varphi,\pi)\in H^{\infty},k=0,1,\cdots \}. $$
Here we state the definition of star products only for single function or functional, the cases of formal power series are defined in the same way as one in section3.
\begin{definition}
For Hamilton densities $f(\partial_{x}^{\alpha}\varphi(x),\partial_{x}^{\beta}\pi(x)),g(\partial_{y}^{\alpha}\varphi(y),\partial_{y}^{\beta}\pi(y))$,
their star product is defined to be

\begin{equation}
\begin{array}{c}
  f(\partial_{x}^{\alpha}\varphi(x),\partial_{x}^{\beta}\pi(x))\star_{P}g(\partial_{y}^{\alpha}\varphi(y),\partial_{y}^{\beta}\pi(y)) \\
  \doteq f(u_{\alpha},\xi_{\beta})\star_{P}g(v_{\alpha},\eta_{\beta})
  |_{u_{\alpha}=\partial_{x}^{\alpha}\varphi(x),\xi_{\beta}=\partial_{x}^{\beta}\pi(x),
  v_{\alpha}=\partial_{y}^{\alpha}\varphi(y),\eta_{\beta}=\partial_{y}^{\beta}\pi(y)}.
\end{array}
\end{equation}
\end{definition}

\begin{definition}
For functional
$$F(\varphi,\pi)=\int_{\mathbb{R}^{n}}f(\partial_{x}^{\alpha}\varphi(x),\partial_{x}^{\beta}\pi(x))dx $$
and Hamilton density $g(\partial_{y}^{\alpha}\varphi(y),\partial_{y}^{\beta}\pi(y))$, we define the following product
\begin{equation}
\begin{array}{c}
  F(\varphi,\pi)\star_{P}g(\partial_{y}^{\alpha}\varphi(y),\partial_{y}^{\beta}\pi(y)) \\
  \doteq\langle f(\partial_{x}^{\alpha}\varphi(x),\partial_{x}^{\beta}\pi(x))\star_{P}g(\partial_{y}^{\alpha}\varphi(y),\partial_{y}^{\beta}\pi(y)) ,1_{x}\rangle.
\end{array}
\end{equation}
Where $f(u_{\alpha},\xi_{\beta})$ satisfies
the condition $\mathbf{B}$.
\end{definition}

\begin{definition}
For two functionals
$$F(\varphi,\pi)=\int_{\mathbb{R}^{n}}f(\partial_{x}^{\alpha}\varphi(x),\partial_{x}^{\beta}\pi(x))dx,$$
$$G(\varphi,\pi)=\int_{\mathbb{R}^{n}}g(\partial_{y}^{\alpha}\varphi(y),\partial_{y}^{\beta}\pi(y))dy, $$
we define their product in the following way
\begin{equation}
\begin{array}{c}
  F(\varphi,\pi)\star_{P}G(\varphi,\pi) \\
  \doteq\int_{\mathbb{R}^{n}}F(\varphi,\pi)\star_{P}g(\partial_{y}^{\alpha}\varphi(y),\partial_{y}^{\beta}\pi(y)) dy.
\end{array}
\end{equation}
where $f(u_{\alpha},\xi_{\beta}),g(v_{\alpha},\eta_{\beta})$ satisfy the condition $\mathbf{B}$.
\end{definition}
Now we have
\begin{prop}
When $P(x,y)=\delta(x-y)$ we have
\begin{equation}
\begin{array}{c}
  F(\varphi,\pi)\star_{\delta}g(\partial_{y}^{\alpha}\varphi(y),\partial_{y}^{\beta}\pi(y))
=F(\varphi,\pi)g(\partial_{y}^{\alpha}\varphi(y),\partial_{y}^{\beta}\pi(y))+ \\
  \{\exp\{\hbar\Gamma\}-1\}(f(\partial_{y}^{\alpha}\varphi(y),\partial_{y}^{\beta}\pi(y))g(v_{\alpha},\eta_{\beta}))
|_{v_{\alpha}=\partial_{y}^{\alpha}\varphi(x),\eta_{\beta}=\partial_{y}^{\beta}\pi(y)},
\end{array}
\end{equation}
where $\Gamma\doteq L_{u,y}\partial_{\eta,y}-L_{\xi,y}\partial_{v,y}$.
\end{prop}
\emph{Proof.}$\;$ Firstly, we note
$$(\partial_{u,x}\partial_{\eta,y})^{i} (\partial_{\xi,x}\partial_{v,y})^{j} (f(u_{\alpha},\xi_{\beta})
g(v_{\alpha},\eta_{\beta})\delta(x-y))|_{u_{\alpha}=\partial_{x}^{\alpha}\varphi(x),\xi_{\beta}=\partial_{x}^{\beta}\pi(x),
v_{\alpha}=\partial_{y}^{\alpha}\varphi(y),\eta_{\beta}=\partial_{y}^{\beta}\pi(y)}$$
$$=\partial_{\eta,y}^{i}\partial_{v,y}^{j}g(v_{\alpha},\eta_{\beta})(\partial_{u,x}^{i}\partial_{\xi,x}^{j}(f,x))\delta(x-y)
|_{v_{\alpha}=\partial_{y}^{\alpha}\varphi(y),\eta_{\beta}=\partial_{y}^{\beta}\pi(y)}.  $$
By the formulas (4.12),(4.13), it is enough for us to check the following case
$$\langle\partial_{\eta,y}^{i}\partial_{v,y}^{j}g(v_{\alpha},\eta_{\beta})(\partial_{u,x}^{i}\partial_{\xi,x}^{j}(f,x))\delta(x-y)
|_{v_{\alpha}=\partial_{y}^{\alpha}\varphi(y),\eta_{\beta}=\partial_{y}^{\beta}\pi(y)} ,1_{x}\rangle   $$
\begin{align*}
&=\partial_{\eta,y}^{i}\partial_{v,y}^{j}g(v_{\alpha},\eta_{\beta})\langle(\partial_{u,x}^{i}\partial_{\xi,x}^{j}(f,x))\delta(x-y),
 1_{x}\rangle|_{v_{\alpha}=\partial_{y}^{\alpha}\varphi(y),\eta_{\beta}=\partial_{y}^{\beta}\pi(y)} \\
&=\partial_{\eta,y}^{i}\partial_{v,y}^{j}g(v_{\alpha},\eta_{\beta})\langle\delta(x-y),(\partial_{u,x}^{i}\partial_{\xi,x}^{j}(f,x))^{t}
 (1_{x})\rangle|_{v_{\alpha}=\partial_{y}^{\alpha}\varphi(y),\eta_{\beta}=\partial_{y}^{\beta}\pi(y)} \\
&=\partial_{\eta,y}^{i}\partial_{v,y}^{j}g(v_{\alpha},\eta_{\beta})\langle\delta(x-y),(L_{u,x})^{i}(L_{\xi,x})^{j}
f(\partial_{x}^{\alpha}\varphi(x),\partial_{x}^{\beta}\pi(x))
\rangle|_{v_{\alpha}=\partial_{y}^{\alpha}\varphi(y),\eta_{\beta}=\partial_{y}^{\beta}\pi(y)}\\
&=\partial_{\eta,y}^{i}\partial_{v,y}^{j}g(v_{\alpha},\eta_{\beta})(L_{u,y})^{i}(L_{\xi,y})^{j}
f(\partial_{y}^{\alpha}\varphi(y),\partial_{y}^{\beta}\pi(y))|_{v_{\alpha}=\partial_{y}^{\alpha}\varphi(y),\eta_{\beta}=\partial_{y}^{\beta}\pi(y)}.
\end{align*}
Thus the formula (4.15) is valid.$\blacksquare$

\begin{prop}
We have
\begin{equation}
\begin{array}{c}
  F(\varphi,\pi)\star_{P}G(\varphi,\pi)=F(\varphi,\pi)G(\varphi,\pi) \\
  +\langle P(x,y),\{\exp\{\hbar\Xi_{\mp}\}-1\}
  f(\partial_{x}^{\alpha}\varphi(x),\partial_{x}^{\beta}\pi(x))g(\partial_{y}^{\alpha}\varphi(y),\partial_{y}^{\beta}\pi(y))\rangle.
\end{array}
\end{equation}
Specially, when $P(x,y)=\delta(x-y)$, we have
\begin{equation}
\begin{array}{c}
  F(\varphi,\pi)\star_{\delta}G(\varphi,\pi)=F(\varphi,\pi)G(\varphi,\pi) \\
  +\int_{\mathbb{R}^{n}}\{\exp\{\hbar\Xi_{-}\}-1\}
  [f(\partial_{x}^{\alpha}\varphi(x),\partial_{x}^{\beta}\pi(x))
  g(\partial_{y}^{\alpha}\varphi(y),\partial_{y}^{\beta}\pi(y))]|_{x=y}dy.
\end{array}
\end{equation}
where $\Xi_{\mp}\doteq L_{u,x}L_{\eta,y}\mp L_{\xi,x}L_{v,y}$.
\end{prop}
\emph{Proof.}$\;$ It is enough for us to check (4.16). At first we note
$$(\partial_{u,x}\partial_{\eta,y})^{i} (\partial_{\xi,x}\partial_{v,y})^{j} (f(u_{\alpha},\xi_{\beta})
g(v_{\alpha},\eta_{\beta})P(x,y)|_{u_{\alpha}=\partial_{x}^{\alpha}\varphi(x),\xi_{\beta}=\partial_{x}^{\beta}\pi(x),
v_{\alpha}=\partial_{y}^{\alpha}\varphi(y),\eta_{\beta}=\partial_{y}^{\beta}\pi(y)}$$
$$=(\partial_{\eta,y}^{i}\partial_{v,y}^{j}(g,y))(\partial_{u,x}^{i}\partial_{\xi,x}^{j}(f,x))P(x,y).  $$
Thus we have
$$\int_{\mathbb{R}^{n}}\langle(\partial_{\eta,y}^{i}\partial_{v,y}^{j}(g,y))(\partial_{u,x}^{i}\partial_{\xi,x}^{j}(f,x))P(x,y),
1_{x}\rangle dy$$
\begin{align*}
 &=\int_{\mathbb{R}^{n}}(\partial_{\eta,y}^{i}\partial_{v,y}^{j}(g,y))\langle(\partial_{u,x}^{i}\partial_{\xi,x}^{j}(f,x))P(x,y),
1_{x}\rangle dy\\
&=\int_{\mathbb{R}^{n}}(\partial_{\eta,y}^{i}\partial_{v,y}^{j}(g,y))\langle P(x,y),
(L_{u,x})^{i}(L_{\xi,x})^{j}f(\partial_{x}^{\alpha}\varphi(x),\partial_{x}^{\beta}\pi(x)) \rangle dy\\
&=\int_{\mathbb{R}^{n}}(L_{\eta,y})^{i}(L_{v,y})^{j}g(\partial_{y}^{\alpha}\varphi(y),\partial_{y}^{\beta}\pi(y))
\langle P(x,y),(L_{u,x})^{i}(L_{\xi,x})^{j}f(\partial_{x}^{\alpha}\varphi(x),\partial_{x}^{\beta}\pi(x)) \rangle dy\\
&=\langle P(x,y),(L_{u,x})^{i}(L_{\xi,x})^{j}f(\partial_{x}^{\alpha}\varphi(x),\partial_{x}^{\beta}\pi(x))
(L_{\eta,y})^{i}(L_{v,y})^{j}g(\partial_{y}^{\alpha}\varphi(y),\partial_{y}^{\beta}\pi(y))\rangle.
\end{align*}
Thus the formula (4.16) is valid. The formula (4.17) is a special case of (4.16).$\blacksquare$

With the help of theorem3.2, we can prove that the associative law is valid for the star products at different levels. Here we state only the conclusion, but omit the proof. It is natural that for three hamiltonian densities
$f(\partial_{x}^{\alpha}\varphi(x),\partial_{x}^{\beta}\pi(x))$,
$g(\partial_{y}^{\alpha}\varphi(y),\partial_{y}^{\beta}\pi(y))$,
$h(\partial_{z}^{\alpha}\varphi(z),\partial_{z}^{\beta}\pi(z))$,
their star product should be defined to be

$$f(\partial_{x}^{\alpha}\varphi(x),\partial_{x}^{\beta}\pi(x))\star_{P}
(g(\partial_{y}^{\alpha}\varphi(y),\partial_{y}^{\beta}\pi(y))\star_{P}h(\partial_{z}^{\alpha}\varphi(z),\partial_{z}^{\beta}\pi(z)))$$
$$\doteq[f(u_{\alpha},\xi_{\beta})\star_{P}(g(v_{\alpha},\eta_{\beta})\star_{P}h(w_{\alpha},\zeta_{\beta}))]
|_{u_{\alpha}=\partial_{x}^{\alpha}\varphi(x),\cdots,\zeta_{\beta}=\partial_{z}^{\beta}\pi(z)},$$
or,
$$(f(\partial_{x}^{\alpha}\varphi(x),\partial_{x}^{\beta}\pi(x))\star_{P}
g(\partial_{y}^{\alpha}\varphi(y),\partial_{y}^{\beta}\pi(y)))\star_{P}
h(\partial_{z}^{\alpha}\varphi(z),\partial_{z}^{\beta}\pi(z))$$
$$\doteq[(f(u_{\alpha},\xi_{\beta})\star_{P}g(v_{\alpha},\eta_{\beta}))\star_{P}h(w_{\alpha},\zeta_{\beta})]
|_{u_{\alpha}=\partial_{x}^{\alpha}\varphi(x),\cdots,\zeta_{\beta}=\partial_{z}^{\beta}\pi(z)}.$$
From theorem3.2 we have
\begin{equation}
\begin{array}{c}
  f(\partial_{x}^{\alpha}\varphi(x),\partial_{x}^{\beta}\pi(x))\star_{P}
(g(\partial_{y}^{\alpha}\varphi(y),\partial_{y}^{\beta}\pi(y))\star_{P}h(\partial_{z}^{\alpha}\varphi(z),\partial_{z}^{\beta}\pi(z))) \\
  =(f(\partial_{x}^{\alpha}\varphi(x),\partial_{x}^{\beta}\pi(x))\star_{P}
g(\partial_{y}^{\alpha}\varphi(y),\partial_{y}^{\beta}\pi(y)))\star_{P}
h(\partial_{z}^{\alpha}\varphi(z),\partial_{z}^{\beta}\pi(z)).
\end{array}
\end{equation}

Now we consider the case of functional. Let
$$F(\varphi,\pi)=\int_{\mathbb{R}^{n}}f(\partial_{x}^{\alpha}\varphi(x),\partial_{x}^{\beta}\pi(x))dx,$$
$$G(\varphi,\pi)=\int_{\mathbb{R}^{n}}g(\partial_{y}^{\alpha}\varphi(y),\partial_{y}^{\beta}\pi(y))dy,$$
$$H(\varphi,\pi)=\int_{\mathbb{R}^{n}}h(\partial_{z}^{\alpha}\varphi(z),\partial_{z}^{\beta}\pi(z))dz.$$
We have

$$F(\varphi,\pi)\star_{P}(g(\partial_{y}^{\alpha}\varphi(y),\partial_{y}^{\beta}\pi(y))\star_{P}
h(\partial_{z}^{\alpha}\varphi(z),\partial_{z}^{\beta}\pi(z)))$$
$$\doteq\langle f(\partial_{x}^{\alpha}\varphi(x),\partial_{x}^{\beta}\pi(x))\star_{P}(g(\partial_{y}^{\alpha}\varphi(y),\partial_{y}^{\beta}\pi(y))\star_{P}
h(\partial_{z}^{\alpha}\varphi(z),\partial_{z}^{\beta}\pi(z))),1_{x}\rangle,$$
and

$$(F(\varphi,\pi)\star_{P}g(\partial_{y}^{\alpha}\varphi(y),\partial_{y}^{\beta}\pi(y)))
  \star_{P}h(\partial_{z}^{\alpha}\varphi(z),\partial_{z}^{\beta}\pi(z))$$
$$\doteq\langle(f(\partial_{x}^{\alpha}\varphi(x),\partial_{x}^{\beta}\pi(x))\star_{P}g(\partial_{y}^{\alpha}\varphi(y),\partial_{y}^{\beta}\pi(y)))   \star_{P}h(\partial_{z}^{\alpha}\varphi(z),\partial_{z}^{\beta}\pi(z)),1_{x}\rangle.$$
Furthermore, we have
\begin{equation}
\begin{array}{c}
  F(\varphi,\pi)\star_{P}(g(\partial_{y}^{\alpha}\varphi(y),\partial_{y}^{\beta}\pi(y))\star_{P}
h(\partial_{z}^{\alpha}\varphi(z),\partial_{z}^{\beta}\pi(z))) \\
  =(F(\varphi,\pi)\star_{P}g(\partial_{y}^{\alpha}\varphi(y),\partial_{y}^{\beta}\pi(y)))
  \star_{P}h(\partial_{z}^{\alpha}\varphi(z),\partial_{z}^{\beta}\pi(z)).
\end{array}
\end{equation}

If we consider the star product of two functionals and one Hamiltonian density we have

$$F(\varphi,\pi)\star_{P}(G(\varphi,\pi)\star_{P}h(\partial_{z}^{\alpha}\varphi(z),\partial_{z}^{\beta}\pi(z)))$$
$$\doteq\langle\langle f\star_{P}(g\star_{P}h),1_{y}\rangle,1_{x}\rangle,$$
$$(F(\varphi,\pi)\star_{P}G(\varphi,\pi))\star_{P}h(\partial_{z}^{\alpha}\varphi(z),\partial_{z}^{\beta}\pi(z))$$
$$\doteq\langle\langle(f\star_{P}g)\star_{P}h,1_{y}\rangle,1_{x}\rangle.$$
It is easy to prove
\begin{equation}
\begin{array}{c}
 F(\varphi,\pi)\star_{P}(G(\varphi,\pi)\star_{P}h(\partial_{z}^{\alpha}\varphi(z),\partial_{z}^{\beta}\pi(z))) \\
  =(F(\varphi,\pi)\star_{P}G(\varphi,\pi))\star_{P}h(\partial_{z}^{\alpha}\varphi(z),\partial_{z}^{\beta}\pi(z)).
\end{array}
\end{equation}
Finally, we have
\begin{equation}
  F(\varphi,\pi)\star_{P}(G(\varphi,\pi)\star_{P}H(\varphi,\pi))
  =(F(\varphi,\pi)\star_{P}G(\varphi,\pi))\star_{P}H(\varphi,\pi).
\end{equation}

\begin{remark}
The star products defined in (4.12),(4.13) and (4.14) are following maps
$$\mathcal{H}_{den,\hbar}\times\mathcal{H}_{den,\hbar}\rightarrow\mathcal{H}_{den,\hbar}\otimes\mathcal{H}_{den,\hbar}\otimes Delta_{x,y},$$
$$H_{\hbar}^{\infty}\times\mathcal{H}_{den,\hbar}\rightarrow H_{\hbar}^{\infty}\otimes\mathcal{H}_{den,\hbar},$$
$$H_{\hbar}^{\infty}\times H_{\hbar}^{\infty}\rightarrow H_{\hbar}^{\infty}.$$

Actually, $H_{\hbar}^{\infty}\otimes\mathcal{H}_{den,\hbar}$ and $H_{\hbar}^{\infty}\otimes\mathcal{H}_{den,\hbar}\otimes Delta_{x,y}$
can be consider as $\mathcal{H}_{den,\hbar}$ modules also, and the star products on them are well defined.
\end{remark}

\begin{remark}
Here we discuss the case of wave equation as an example of physics. The Lagrangian density is of form as following
$$\mathcal{L}=\mathcal{L}_{Klein-Gordon}+\mathcal{L}_{int}=\frac{1}{2}[\pi^{2}-|\nabla_{x}\varphi|^{2}-m^{2}\varphi^{2}]-U(\varphi), $$
where $\varphi$ is a smooth map from $\mathbb{R}_{t}$ to $\mathcal{S}(\mathbb{R}_{x}^{3})$, $U(s)$ is a smooth function,
$U(s)=o(s^{2}),(s\rightarrow0)$, and $\dot{\varphi}=\frac{\partial\varphi}{\partial t}$, $\nabla_{x}\varphi
=\{\varphi_{x_{1}},\varphi_{x_{2}},\varphi_{x_{3}}\}$. The Hamiltonian density is
$$\mathcal{H}(\varphi,\nabla\varphi,\pi)=\pi\frac{\partial\mathcal{L}}{\partial\dot{\varphi}}-\mathcal{L}=\frac{1}{2}
[\pi^{2}+|\nabla_{x}\varphi|^{2}+m^{2}\varphi^{2}]+U(\varphi),$$
where $\pi=\frac{\partial\mathcal{L}}{\partial\dot{\varphi}}=\dot{\varphi}$. The basic Poisson brackets are
$$\{\pi(t,x),\varphi(t,y)\}=-i\delta(x-y),\{\pi(t,x),\pi(t,y)\}=0,\{\varphi(t,x),\varphi(t,y)\}=0.$$
The Hamiltonian functional is
$$H(\varphi,\pi)=\int\mathcal{H}(\varphi,\nabla\varphi,\pi)dx.$$
Then the equation of motion is
$$\dot{\pi}=i(H(\varphi,\pi)\star_{i\delta}\pi-\pi\star_{i\delta}H(\varphi,\pi))=i\{H(\varphi,\pi),\pi\}_{i\delta}$$
$$=\triangle_{x}\varphi-m^{2}\varphi-U^{\prime}(\varphi),$$
where $\triangle_{x}=\frac{\partial^{2}}{\partial x_{1}^{2}}+\frac{\partial^{2}}{\partial x_{2}^{2}}+\frac{\partial^{2}}{\partial x_{3}^{2}}$,
and we take $\hbar=1$.
\end{remark}

\subsection{The star product related to Peierls bracket}

Let $F\in \mathcal{S'}(\mathbb{R}^{n})$ be a temperate distribution, then convolution operator
$$F\ast :\varphi \mapsto F\ast \varphi, \varphi \in \mathcal{S}(\mathbb{R}^{n})$$
is a smooth linear map from $\mathcal{S}(\mathbb{R}^{n})$ to $C^{\infty}(\mathbb{R}^{n})$, where $\mathcal{S}(\mathbb{R}^{n})$ and $C^{\infty}(\mathbb{R}^{n})$ are endowed with natural topology. It is easy to check that we have
$$\frac{\delta (F\ast \varphi)(x)}{\delta \varphi(y)}=F(x-y).$$
Actually, $$\frac{d}{dt}F\ast (\varphi+t\delta\varphi)|_{t=0}=F\ast \delta\varphi=\langle F(x-y),\delta\varphi(y)\rangle.$$
On the other hand, from definition4.1 and formula(2.14) we know that
$$\{f(\partial _{x}^{\alpha}\varphi(x),\partial _{x}^{\beta}\pi(x)),\pi(y)\}=\frac{\delta f(\partial _{x}^{\alpha}\varphi(x),\partial _{x}^{\beta}\pi(x))}{\delta \varphi(y)}.$$
Above formulas suggest us an idea to generalize  Poisson bracket in the following way
$$\{(F\ast \varphi)(x),\pi(y)\}\doteq F\ast \{\varphi(x),\pi(y)\}_{\delta}.$$

To discuss Peierls bracket, we work on $\mathbb{R}^{n+1}=\mathbb{R}_{t}\oplus\mathbb{R}_{x}^{n}$ endowed with Lorentzian metric $dt\otimes dt-\sum_{i=1}^{n}dx_{i}\otimes dx_{i}$  and consider Cauchy problem of Klein-Gordon equation as following

\begin{equation}
  \left \{\begin{array}{c}
(\partial_{t}^{2}-\triangle_{x}+m^{2})\phi(t,x)=0, \\
\phi(0,x)=\varphi(x),\phi_{t}(0,x)=\pi(x),
  \end{array}
 \right.
\end{equation}
where $\varphi(x),\pi(x)\in \mathcal{S}(\mathbb{R}^{n})$. The solutions of Cauchy problem (4.22) are given by
\begin{equation}
  \phi(t,x)=(\triangle\ast\pi)(x)+(\frac{\partial\triangle}{\partial t}\ast\varphi)(x),
\end{equation}
where $\triangle(t,x)$ is Green function satisfying
$$
\left \{\begin{array}{c}
(\partial_{t}^{2}-\triangle_{x}+m^{2})\triangle(t,x)=0, \\
\triangle(0,x)=\delta(x),\frac{\partial\triangle}{\partial t}(0,x)=0.
  \end{array}
 \right.
$$

Now we define Poisson bracket as following
\begin{equation}
  \{\phi(t,x),\varphi(y)\}\doteq\triangle\ast\{\pi(x),\varphi(y)\}_{\delta}+\frac{\partial\triangle}{\partial t}\ast\{\varphi(x),\varphi(y)\}_{\delta}.
\end{equation}
Hence we have
\begin{equation}
  \{\phi(t,x),\varphi(y)\}=-\triangle(t,x-y).
\end{equation}
From Poisson bracket (4.25) we can discuss the Poisson bracket between $\phi(t,x)$ and $\phi(s,y)$, without loss of generality we suppose
$t>s$, due to the uniqueness of Cauchy problem (4.22) we have
$$\phi(t,x)=(\triangle(t-s,\cdot)\ast\frac{\partial\phi(t,\cdot)}{\partial t}|_{t=s})(x)+
(\frac{\partial\triangle(t-s,\cdot)}{\partial t}\ast\phi(s,\cdot))(x).$$
With the help of (4.25) we have
\begin{equation}
  \{\phi(t,x),\phi(s,y)\}=-\triangle(t-s,x-y).
\end{equation}
The formula (4.26) just be Peierls bracket at level of fields.

Furthermore, we define the star product related to Peierls bracket as following

\begin{equation}
  \begin{array}{c}
    \phi(t,x)\star\phi(s,y)\doteq\triangle(t-s,\cdot)\ast(\phi_{t}(s,x)\star_{\delta}\phi(s,y))\\
    +\frac{\partial\triangle(t-s,\cdot)}{\partial t}\ast(\phi(s,x)\star_{\delta}\phi(s,y)).
  \end{array}
\end{equation}
By a straightforward calculation we get the following formula
\begin{equation}
  \phi(t,x)\star\phi(s,y)=\phi(t,x)\cdot\phi(s,y)+\hbar\{\phi(t,x),\phi(s,y)\}.
\end{equation}
The formula (4.28) is same as one of star product in R. Brunetti, M. Dutsch, K. Fredenhagen\cite{g}(formula(15) in \cite{g}).
The formulas (4.24) and (4.27) show that the construction under the Hamiltonian formulation plays the role of Cauchy data concerning
the star product on $\mathbb{R}^{n+1}$.

\section{The case of complex scalar fields}

In this section the all discussion is parallel to the case of real scalar fields which is discussed from section2 to section4, thus we state only the definition and conclusion omitting the proof at all.

\subsection{Notations}

 We consider Hamiltonian functions in the case of complex scalar fields as smooth functions $f(z_{\alpha},\bar{z}_{\beta})$,
where the variables of the function are finite subsets of $\{(z_{\alpha},\bar{z}_{\beta})|z_{\alpha},\bar{z}_{\beta}\in\mathbb{C},
\alpha,\beta\in\mathbb{N}^{n}\}$. The space of Hamiltonian functions in the case of complex scalar fields is denoted by
$\mathcal{H}_{\mathbb{C}}^{\infty}$. The Hamiltonian densities are functions
$$f(\partial_{x}^{\alpha}\psi(x),\partial_{x}^{\beta}\bar{\psi}(x))=f(z_{\alpha},\bar{z}_{\beta})
|_{z_{\alpha}=\partial_{x}^{\alpha}\psi(x),\bar{z}_{\beta}=\partial_{x}^{\beta}\bar{\psi}(x)},\psi(x)\in\mathcal{S}(\mathbb{R}^{n}).$$
The set of all Hamiltonian densities is denoted by $\mathcal{H}_{\mathbb{C},den}$. The Hamiltonian functional are of form as following
$$F(\psi,\bar{\psi})=\int_{\mathbb{R}^{n}}f(\partial_{x}^{\alpha}\psi(x),\partial_{x}^{\beta}\bar{\psi}(x))dx,\psi(x)\in\mathcal{S}(\mathbb{R}^{n}).$$
On the other hand we need to introduce the condition similar to the condition $\mathbf{B}$ in section2 as following
\begin{equation}
  f(z_{\alpha},\bar{z}_{\beta})|_{z_{\alpha}=0,\bar{z}_{\beta}=0}.
\end{equation}
We call (5.1) condition $\mathbf{B^{\ast}}$. We assume that Hamiltonian functions concerning the Hamiltonian functional satisfy the condition
$\mathbf{B^{\ast}}$. The set of Hamiltonian functionals for complex scalar fields is denoted by $H_{\mathbb{C}}^{\infty}$.

Similar to the discussion in section2, we can define the algebras of the Euler-Lagrange operators and the dual Euler-Lagrange operators
denoted by $\mathcal{L}_{\mathbb{C}}$ and $\mathcal{L}_{\mathbb{C}}^{t}$ respectively. Specially, for Euler-Lagrange derivatives in the case of complex fields we have
\begin{equation}
  \partial_{z,x}\doteq\sum_{\alpha\in\mathbb{N}^{n}}\frac{\partial}{\partial z_{\alpha}}\partial_{x}^{\alpha},
\partial_{\bar{z},x}\doteq\sum_{\beta\in\mathbb{N}^{n}}\frac{\partial}{\partial \bar{z}_{\beta}}\partial_{x}^{\beta}.
\end{equation}
Where $\partial_{z,x},\partial_{\bar{z},x}$ are maps as followings
$$\mathcal{H}_{\mathbb{C}}^{\infty}\times Delta_{x,y}\rightarrow\mathcal{H}_{\mathbb{C}}^{\infty}\otimes Delta_{x,y},$$
$$\mathcal{H}_{\mathbb{C}}^{\infty}\times C^{\infty}(\mathbb{R}_{x}^{n})\rightarrow
\mathcal{H}_{\mathbb{C}}^{\infty}\otimes C^{\infty}(\mathbb{R}_{x}^{n}),$$
$$\mathcal{H}_{\mathbb{C},den}\times Delta_{x,y}\rightarrow
\mathcal{H}_{\mathbb{C},den}\otimes Delta_{x,y},$$
$$\mathcal{H}_{\mathbb{C},den}\times C^{\infty}(\mathbb{R}_{x}^{n})\rightarrow
\mathcal{H}_{\mathbb{C},den}\otimes C^{\infty}(\mathbb{R}_{x}^{n}).$$
Furthermore, for a Hamiltonian density $f(\partial_{x}^{\alpha}\psi(x),\partial_{x}^{\beta}\bar{\psi}(x))$ the related Euler-Lagrange operators $\partial_{z,x}(f,x),\partial_{\bar{z},x}(f,x)$ can be defined in the same way as section2.

The dual Euler-Lagrange derivatives can be defined also, they are
\begin{equation}
  L_{z,x}\doteq\sum_{\alpha\in\mathbb{N}^{n}}(-1)^{|\alpha|}\partial_{x}^{\alpha}\frac{\partial}{\partial z_{\alpha}},
  L_{\bar{z},x}\doteq\sum_{\beta\in\mathbb{N}^{n}}(-1)^{|\beta|}\partial_{x}^{\beta}\frac{\partial}{\partial z_{\beta}},
\end{equation}
and then for a Hamiltonian density $f(\partial_{x}^{\alpha}\psi(x),\partial_{x}^{\beta}\bar{\psi}(x))$ we have
\begin{equation}
  [\partial_{z,x}(f,x)]^{t}(1_{x})=L_{z,x}f(\cdot),[\partial_{\bar{z},x}(f,x)]^{t}(1_{x})=L_{\bar{z},x}f(\cdot).
\end{equation}

\subsection{The Poisson brackets}

We define the Poisson brackets in the same way as one in previous sections.
\begin{definition}
Let $P(x,y)\in Delta_{x,y}$, for two Hamilton functions $f(z_{\alpha},\bar{z}_{\beta})$, $g(w_{\alpha},\bar{w}_{\beta})\in\mathcal{H}_{\mathbb{C}}^{\infty}$, we assign $f(z_{\alpha},\bar{z}_{\beta})$, $g(w_{\alpha},\bar{w}_{\beta})$
to the variables $x,y\in\mathbb{R}^{n}$ respectively, their Poisson bracket is defined in the following way.
\begin{equation}
\begin{array}{c}
  \{f(z_{\alpha},\bar{z}_{\beta}),g(w_{\alpha},\bar{w}_{\beta})\}_{P} \\
  \doteq(\partial_{z,x}\partial_{\bar{w},y}\mp\partial_{\bar{z},x}\partial_{w,y})
  f(z_{\alpha},\bar{z}_{\beta})g(w_{\alpha},\bar{w}_{\beta})P(x,y).
\end{array}
\end{equation}
Where minus sign in (5.5) corresponds to the symmetric distribution P(x,y), and plus sign corresponds to the anti-symmetric case.
\end{definition}

\begin{definition}
For two Hamilton densities $f(\partial_{x}^{\alpha}\psi(x),\partial_{x}^{\beta}\bar{\psi}(x))$,
$g(\partial_{y}^{\alpha}\psi(y),\partial_{y}^{\beta}\bar{\psi}(y))$ $\in\mathcal{H}_{\mathbb{C},den}$, their Poisson bracket is defined to be
\begin{equation}
\begin{array}{c}
  \{f(\partial_{x}^{\alpha}\psi(x),\partial_{x}^{\beta}\bar{\psi}(x)),
g(\partial_{y}^{\alpha}\psi(y),\partial_{y}^{\beta}\bar{\psi}(y))\}_{P} \\
   \doteq\{f(z_{\alpha},\bar{z}_{\beta}),g(w_{\alpha},\bar{w}_{\beta})\}_{P}
  |_{z_{\alpha}=\partial_{x}^{\alpha}\psi(x),\bar{z}_{\beta}=\partial_{x}^{\beta}\bar{\psi}(x),
  w_{\alpha}=\partial_{y}^{\alpha}\psi(y),\bar{w}_{\beta}=\partial_{y}^{\beta}\bar{\psi}(y)}.
\end{array}
\end{equation}
\end{definition}

\begin{definition}
For $F(\psi,\bar{\psi}),G(\psi,\bar{\psi})\in H_{\mathbb{C}}^{\infty}$, $f(\partial_{x}^{\alpha}\psi(x),\partial_{x}^{\beta}\bar{\psi}(x))$,
$g(\partial_{y}^{\alpha}\psi(y),\partial_{y}^{\beta}\bar{\psi}(y))$ $\in\mathcal{H}_{\mathbb{C},den}$, we define the following Poisson
brackets
\begin{equation}
  \{F(\psi,\bar{\psi}),g(\partial_{y}^{\alpha}\psi(y),\partial_{y}^{\beta}\bar{\psi}(y))\}_{P}
  \doteq\langle\{f,g\}_{P},1_{x}\rangle,
\end{equation}
\begin{equation}
  \{F(\psi,\bar{\psi}),G(\psi,\bar{\psi})\}_{P}\doteq \int_{\mathbb{R}^{n}}\langle\{f,g\}_{P},1_{x}\rangle dy.
\end{equation}
\end{definition}

Similar to case of real scalar fields, $H_{\mathbb{C}}^{\infty}\otimes\mathcal{H}_{\mathbb{C},den}$ and
$H_{\mathbb{C}}^{\infty}\otimes\mathcal{H}_{\mathbb{C},den}\otimes Delta_{x,y}$ can be considered as
$\mathcal{H}_{\mathbb{C},den}$ modules, so we need the following definition
\begin{definition}
For $F(\psi,\bar{\psi}),G(\psi,\bar{\psi})\in H_{\mathbb{C}}^{\infty}$, $f(\partial_{x}^{\alpha}\psi(x),\partial_{x}^{\beta}\bar{\psi}(x))$,
$g(\partial_{y}^{\alpha}\psi(y),\partial_{y}^{\beta}\bar{\psi}(y))$, $h(\partial_{z}^{\alpha}\psi(z),\partial_{z}^{\beta}\bar{\psi}(z))$
$\in\mathcal{H}_{\mathbb{C},den}$, we have
\begin{equation}
  \{F,Gh\}_{P}\doteq G\{F,h\}_{P}+h\{F,G\}_{P},
\end{equation}
\begin{equation}
  \{f,Gh\}_{P}\doteq G\{f,h\}_{P}+h\{f,G\}_{P},
\end{equation}
where
$$F(\psi,\bar{\psi})=\int_{\mathbb{R}^{n}}f(\partial_{x}^{\alpha}\psi(x),\partial_{x}^{\beta}\bar{\psi}(x))dx,\;
G(\psi,\bar{\psi})=\int_{\mathbb{R}^{n}}g(\partial_{x}^{\alpha}\psi(x),\partial_{x}^{\beta}\bar{\psi}(x))dx.$$
\end{definition}
\begin{remark}
Let $\varphi(x)=\frac{1}{2}[\psi(x)+\bar{\psi}(x)]$, $\pi(x)=\frac{1}{2i}[\psi(x)-\bar{\psi}(x)]$, it is easy to check that the system of Poisson brackets
$$\{\varphi(x),\pi(y)\}_{P}=P(x,y),\{\varphi(x),\varphi(y)\}_{P}=0,\{\pi(x),\pi(y)\}_{P}=0$$
is equivalent to the following Poisson brackets
$$\{\psi(x),\bar{\psi}(y)\}_{-2iP}=-2iP(x,y),\{\psi(x),\psi(y)\}_{-2iP}=0,\{\bar{\psi}(x),\bar{\psi}(y)\}_{-2iP}=0.$$
\end{remark}
\begin{remark}
It is obvious that the Poisson brackets in (5.5),(5.6),(5.7) and (5.8) are anti-symmetric, bi-linear and derivative for two variables.
\end{remark}
\begin{remark}
Similar to the discussion in subsection3.2 we can discuss the multiple Poisson brackets in the case of complex scalar fields and prove that the Jacobi identity is valid.
\end{remark}

\subsection{The star products}

Actually, due to the discussion in previous sections we know the star products should be defined for formal power series with coefficients being Hamiltonian functions, densities or functionals. However, from technical viewpoint, it is enough for us to discuss the star products for Hamiltonian functions, densities or functionals themselves.

We introduce a notation same as section3 as following
\begin{equation}
\exp(\hbar\sigma_{z,w;x,y}^{\mp})=\sum_{k\geq 0}\frac{\hbar^{k}}{k!}[\sigma_{z,w;x,y}^{\mp}]^{k},
\end{equation}
where
$$
\begin{array}{c}
  [\sigma_{z,w;x,y}^{\mp}]^{k}(f(z_{\alpha},\bar{z}_{\beta}),g(w_{\alpha},\bar{w}_{\beta})) \\
  \doteq(\partial_{z,x}\partial_{\bar{w},y}\mp\partial_{\bar{z},x}\partial_{w,y})^{k}
  f(z_{\alpha},\bar{z}_{\beta})g(w_{\alpha},\bar{w}_{\beta})P(x,y),
\end{array}
$$
$k\geq1$, and
$$
[\sigma_{z,w;x,y}^{\mp}]^{0}(f(z_{\alpha},\bar{z}_{\beta}),g(w_{\alpha},\bar{w}_{\beta}))
  \doteq f(z_{\alpha},\bar{z}_{\beta})g(w_{\alpha},\bar{w}_{\beta}),
$$
then we have
\begin{definition}
Let $P(x,y)\in Delta_{x,y}$, $f(z_{\alpha},\bar{z}_{\beta}),g(w_{\alpha},\bar{w}_{\beta})\in\mathcal{H}_{\mathbb{C},den}$, where
$f(z_{\alpha},\bar{z}_{\beta}),g(w_{\alpha},\bar{w}_{\beta})$ are assigned to the variables $x,y\in\mathbb{R}^{n}$ respectively,
we define
\begin{equation}
  f(z_{\alpha},\bar{z}_{\beta})\star_{P}g(w_{\alpha},\bar{w}_{\beta})
  \doteq\exp(\hbar\sigma_{z,w;x,y}^{\mp})f(z_{\alpha},\bar{z}_{\beta})g(w_{\alpha},\bar{w}_{\beta}),
\end{equation}
For two Hamiltonian densities $f(\partial_{x}^{\alpha}\psi(x),\partial_{x}^{\beta}\bar{\psi}(x))$,
$g(\partial_{y}^{\alpha}\psi(y),\partial_{y}^{\beta}\bar{\psi}(y))$ we define

\begin{equation}
\begin{array}{c}
  f(\partial_{x}^{\alpha}\psi(x),\partial_{x}^{\beta}\bar{\psi}(x))\star_{P}g(\partial_{y}^{\alpha}\psi(y),\partial_{y}^{\beta}\bar{\psi}(y)) \\
  \doteq f(z_{\alpha},\bar{z}_{\beta})\star_{P}g(w_{\alpha},\bar{w}_{\beta})
  |_{z_{\alpha}=\partial_{x}^{\alpha}\psi(x),\bar{z}_{\beta}=\partial_{x}^{\beta}\bar{\psi}(x),
  w_{\alpha}=\partial_{y}^{\alpha}\psi(y),\bar{w}_{\beta}=\partial_{y}^{\beta}\bar{\psi}(y)}.
\end{array}
\end{equation}
\end{definition}
\begin{definition}
For $F(\psi,\bar{\psi}),G(\psi,\bar{\psi})\in H_{\mathbb{C}}^{\infty}$, $f(\partial_{x}^{\alpha}\psi(x),\partial_{x}^{\beta}\bar{\psi}(x)),
g(\partial_{y}^{\alpha}\psi(y),\partial_{y}^{\beta}\bar{\psi}(y))$ $\in\mathcal{H}_{\mathbb{C},den}$, their star products
are defined to be
$$F(\psi,\bar{\psi})\star_{P}g(\partial_{y}^{\alpha}\psi(y),\partial_{y}^{\beta}\bar{\psi}(y))$$
\begin{equation}
  \doteq\langle f(\partial_{x}^{\alpha}\psi(x),\partial_{x}^{\beta}\bar{\psi}(x))\star_{P}
  g(\partial_{y}^{\alpha}\psi(y),\partial_{y}^{\beta}\bar{\psi}(y)),1_{x} \rangle,
\end{equation}
$$F(\psi,\bar{\psi})\star_{P}G(\psi,\bar{\psi})$$
\begin{equation}
  \doteq\int_{\mathbb{R}_{y}^{n}}\langle f(\partial_{x}^{\alpha}\psi(x),\partial_{x}^{\beta}\bar{\psi}(x))\star_{P}
  g(\partial_{y}^{\alpha}\psi(y),\partial_{y}^{\beta}\bar{\psi}(y)),1_{x} \rangle dy.
\end{equation}
Where
$$F(\psi,\bar{\psi})=\int_{\mathbb{R}^{n}}f(\partial_{x}^{\alpha}\psi(x),\partial_{x}^{\beta}\bar{\psi}(x))dx,$$
$$G(\psi,\bar{\psi})=\int_{\mathbb{R}^{n}}g(\partial_{x}^{\alpha}\psi(x),\partial_{x}^{\beta}\bar{\psi}(x))dx,$$
and $f(\partial_{x}^{\alpha}\psi(x),\partial_{x}^{\beta}\bar{\psi}(x))$,
$ g(\partial_{y}^{\alpha}\psi(y),\partial_{y}^{\beta}\bar{\psi}(y))$ satisfy the condition $\mathbf{B^{\ast}}$.
\end{definition}

\begin{propsition}
When $P(x,y)=\delta(x-y)$ we have
\begin{equation}
\begin{array}{c}
  F(\psi,\bar{\psi})\star_{\delta}g(\partial_{y}^{\alpha}\psi(y),\partial_{y}^{\beta}\bar{\psi}(y))
=F(\psi,\bar{\psi})g(\partial_{y}^{\alpha}\psi(y),\partial_{y}^{\beta}\bar{\psi}(y))+ \\
  \{\exp\{\hbar\Gamma_{\mathbb{C}}\}-1\}(g(v_{\alpha},\bar{v}_{\beta})f(\partial_{y}^{\alpha}\psi(y),\partial_{y}^{\beta}\bar{\psi}(y)))
  |_{v_{\alpha}=\partial_{y}^{\alpha}\psi(y),\bar{v}_{\beta}=\partial_{y}^{\beta}\bar{\psi}(y)}.
\end{array}
\end{equation}
Where $\Gamma_{\mathbb{C}}=L_{z,y}\partial_{\bar{v},y}-L_{\bar{z},y}\partial_{v,y}$.
\end{propsition}

\begin{propsition}
We have
\begin{equation}
\begin{array}{c}
  F(\psi,\bar{\psi})\star_{P}G(\psi,\bar{\psi})=F(\psi,\bar{\psi})G(\psi,\bar{\psi})+ \\
  \langle P(x,y),\{\exp\{\hbar\Xi_{\mathbb{C},\mp}\}-1\}(f(\partial_{x}^{\alpha}\psi(x),\partial_{x}^{\beta}\bar{\psi}(x))
  g(\partial_{y}^{\alpha}\psi(y),\partial_{y}^{\beta}\bar{\psi}(y)))\rangle.
\end{array}
\end{equation}
When $P(x,y)=\delta(x-y)$ we have
\begin{equation}
\begin{array}{c}
  F(\psi,\bar{\psi})\star_{\delta}G(\psi,\bar{\psi})=F(\psi,\bar{\psi})G(\psi,\bar{\psi})+ \\
  \int_{\mathbb{R}^{n}}\{\exp\{\hbar\Xi_{\mathbb{C},-}\}-1\}(f(\partial_{x}^{\alpha}\psi(x),\partial_{x}^{\beta}\bar{\psi}(x))
  g(\partial_{y}^{\alpha}\psi(y),\partial_{y}^{\beta}\bar{\psi}(y)))|_{x=y}dy.
\end{array}
\end{equation}
Where $\Xi_{\mathbb{C},\mp}=L_{z,x}L_{\bar{v},y}\mp L_{\bar{z},y}L_{v,y}$.
\end{propsition}

Now we can define the star product among three Hamiltonian functions $f(u_{\alpha},\bar{u}_{\beta}),g(v_{\alpha},\bar{v}_{\beta}),
h(w_{\alpha},\bar{w}_{\beta})$ in the following ways

\begin{equation}
\begin{array}{c}
  f(u_{\alpha},\bar{u}_{\beta})\star_{P}(g(v_{\alpha},\bar{v}_{\beta})h(w_{\alpha},\bar{w}_{\beta})) \\
  \doteq\exp(\hbar\sigma_{u,v;x,y}^{\mp})\exp(\hbar\sigma_{u,w;x,z}^{\mp})(f(\cdot)g(\cdot)h(\cdot)),
\end{array}
\end{equation}
and,
\begin{equation}
\begin{array}{c}
  (f(u_{\alpha},\bar{u}_{\beta})g(v_{\alpha},\bar{v}_{\beta}))\star_{P}h(w_{\alpha},\bar{w}_{\beta})) \\
  \doteq \exp(\hbar\sigma_{u,w;x,z}^{\mp})\exp(\hbar\sigma_{v,w;y,z}^{\mp})(f(\cdot)g(\cdot)h(\cdot)),
\end{array}
\end{equation}
Where we assign $f(u_{\alpha},\bar{u}_{\beta}),g(v_{\alpha},\bar{v}_{\beta}),
h(w_{\alpha},\bar{w}_{\beta})$ to the variables $x,y,z\in\mathbb{R}^{n}$.

\begin{propsition}
For star products in (5.19), (5.20) we have
$$\exp(\hbar\sigma_{u,v;x,y}^{\mp})[\exp(\hbar\sigma_{u,w;x,z}^{\mp})(f(\cdot)g(\cdot)h(\cdot))]$$
$$=\exp(\hbar\sigma_{u,w;x,z}^{\mp})[\exp(\hbar\sigma_{u,v;x,y}^{\mp})(f(\cdot)g(\cdot)h(\cdot))],$$
and
$$\exp(\hbar\sigma_{u,w;x,z}^{\mp})[\exp(\hbar\sigma_{v,w;y,z}^{\mp})(f(\cdot)g(\cdot)h(\cdot))]$$
$$=\exp(\hbar\sigma_{v,w;y,z}^{\mp})[\exp(\hbar\sigma_{u,w;x,z}^{\mp})(f(\cdot)g(\cdot)h(\cdot))].$$
\end{propsition}

The proposition5.1 means that the star products in (5.19), (5.20) are well defined.

With the help of the star products in (5.19), (5.20) we can define the following star product
$f(u_{\alpha},\bar{u}_{\beta}),g(v_{\alpha},\bar{v}_{\beta}),
h(w_{\alpha},\bar{w}_{\beta})$
$$f(u_{\alpha},\bar{u}_{\beta})\star_{P}g(v_{\alpha},\bar{v}_{\beta})\star_{P}h(w_{\alpha},\bar{w}_{\beta})$$
$$\doteq f(u_{\alpha},\bar{u}_{\beta})\star_{P}(g(v_{\alpha},\bar{v}_{\beta})\star_{P}h(w_{\alpha},\bar{w}_{\beta})),$$
or,
$$f(u_{\alpha},\bar{u}_{\beta})\star_{P}g(v_{\alpha},\bar{v}_{\beta})\star_{P}h(w_{\alpha},\bar{w}_{\beta})$$
$$\doteq(f(u_{\alpha},\bar{u}_{\beta})\star_{P}g(v_{\alpha},\bar{v}_{\beta}))\star_{P}h(w_{\alpha},\bar{w}_{\beta}).$$

\begin{theorem}
\begin{equation}
\begin{array}{c}
  f(u_{\alpha},\bar{u}_{\beta})\star_{P}(g(v_{\alpha},\bar{v}_{\beta})\star_{P}h(w_{\alpha},\bar{w}_{\beta})) \\
  =(f(u_{\alpha},\bar{u}_{\beta})\star_{P}g(v_{\alpha},\bar{v}_{\beta}))\star_{P}h(w_{\alpha},\bar{w}_{\beta}).
\end{array}
\end{equation}
\end{theorem}
Theorem5.1 means that the star products defined above satisfy the associative law.

For three Hamiltonian densities$\,$ $f(\partial_{x}^{\alpha}\psi(x),\partial_{x}^{\beta}\bar{\psi}(x))$,$\;$
$g(\partial_{y}^{\alpha}\psi(y),\partial_{y}^{\beta}\bar{\psi}(y))$, $h(\partial_{z}^{\alpha}\psi(z),\partial_{z}^{\beta}\bar{\psi}(z))$
we have
$$f(\partial_{x}^{\alpha}\psi(x),\partial_{x}^{\beta}\bar{\psi}(x))\star_{P}
(g(\partial_{y}^{\alpha}\psi(y),\partial_{y}^{\beta}\bar{\psi}(y))\star_{P}
h(\partial_{z}^{\alpha}\psi(z),\partial_{z}^{\beta}\bar{\psi}(z)))$$
$$\doteq f(u_{\alpha},\bar{u}_{\beta})\star_{P}(g(v_{\alpha},\bar{v}_{\beta})\star_{P}h(w_{\alpha},\bar{w}_{\beta}))
|_{u_{\alpha}=\partial_{x}^{\alpha}\psi(x),\cdots,\bar{w}_{\beta}=\partial_{x}^{\beta}\bar{\psi}(z)},$$
or,
$$(f(\partial_{x}^{\alpha}\psi(x),\partial_{x}^{\beta}\bar{\psi}(x)))\star_{P}
g(\partial_{y}^{\alpha}\psi(y),\partial_{y}^{\beta}\bar{\psi}(y)))\star_{P}
h(\partial_{z}^{\alpha}\psi(z),\partial_{z}^{\beta}\bar{\psi}(z))$$
$$\doteq(f(u_{\alpha},\bar{u}_{\beta})\star_{P}g(v_{\alpha},\bar{v}_{\beta}))\star_{P}h(w_{\alpha},\bar{w}_{\beta})
|_{u_{\alpha}=\partial_{x}^{\alpha}\psi(x),\cdots,\bar{w}_{\beta}=\partial_{x}^{\beta}\bar{\psi}(z)},$$
therefore
$$f(\partial_{x}^{\alpha}\psi(x),\partial_{x}^{\beta}\bar{\psi}(x))\star_{P}
(g(\partial_{y}^{\alpha}\psi(y),\partial_{y}^{\beta}\bar{\psi}(y))\star_{P}
h(\partial_{z}^{\alpha}\psi(z),\partial_{z}^{\beta}\bar{\psi}(z)))$$
$$=(f(\partial_{x}^{\alpha}\psi(x),\partial_{x}^{\beta}\bar{\psi}(x)))\star_{P}
g(\partial_{y}^{\alpha}\psi(y),\partial_{y}^{\beta}\bar{\psi}(y)))\star_{P}
h(\partial_{z}^{\alpha}\psi(z),\partial_{z}^{\beta}\bar{\psi}(z)).$$

Let
$$F(\psi,\bar{\psi})=\int_{\mathbb{R}^{n}}f(\partial_{x}^{\alpha}\psi(x),\partial_{x}^{\beta}\bar{\psi}(x))dx,$$
$$G(\psi,\bar{\psi})=\int_{\mathbb{R}^{n}}g(\partial_{y}^{\alpha}\psi(y),\partial_{y}^{\beta}\bar{\psi}(y))dy,$$
$$H(\psi,\bar{\psi})=\int_{\mathbb{R}^{n}}h(\partial_{z}^{\alpha}\psi(z),\partial_{z}^{\beta}\bar{\psi}(z))dz.$$
We define
$$F(\psi,\bar{\psi})\star_{P}g(\partial_{y}^{\alpha}\psi(y),\partial_{y}^{\beta}\bar{\psi}(y))
\star_{P}h(\partial_{z}^{\alpha}\psi(z),\partial_{z}^{\beta}\bar{\psi}(z))$$

$$\doteq\langle f(\partial_{x}^{\alpha}\psi(x),\partial_{x}^{\beta}\bar{\psi}(x))\star_{P}
g(\partial_{y}^{\alpha}\psi(y),\partial_{y}^{\beta}\bar{\psi}(y))\star_{P}
h(\partial_{z}^{\alpha}\psi(z),\partial_{z}^{\beta}\bar{\psi}(z)),1_{x}\rangle,$$

$$F(\psi,\bar{\psi})\star_{P}G(\psi,\bar{\psi})\star_{P}h(\partial_{z}^{\alpha}\psi(z),\partial_{z}^{\beta}\bar{\psi}(z))$$

$$\doteq\langle\langle f(\partial_{x}^{\alpha}\psi(x),\partial_{x}^{\beta}\bar{\psi}(x))\star_{P}
g(\partial_{y}^{\alpha}\psi(y),\partial_{y}^{\beta}\bar{\psi}(y))\star_{P}
h(\partial_{z}^{\alpha}\psi(z),\partial_{z}^{\beta}\bar{\psi}(z)),1_{x} \rangle,1_{y} \rangle,$$

$$F(\psi,\bar{\psi})\star_{P}G(\psi,\bar{\psi})\star_{P}H(\psi,\bar{\psi})$$

$$\doteq\int_{\mathbb{R}^{n}}\langle\langle f\star_{P}g\star_{P}h,1_{x}\rangle,1_{y}\rangle dz.$$

From above discussion we know that the star products on $\mathcal{H}_{\mathbb{C},den}$, $H_{\mathbb{C}}^{\infty}$, and
$H_{\mathbb{C}}^{\infty}\otimes\mathcal{H}_{\mathbb{C},den}$ satisfy the associative law where
$H_{\mathbb{C}}^{\infty}\otimes\mathcal{H}_{\mathbb{C},den}$ is considered as a $\mathcal{H}_{\mathbb{C},den}$ module.
\begin{remark}
Here we consider nonlinear Schrodinger equation as an example of complex scalar fields. The Hamiltonian functional is
$$H(\psi,\bar{\psi})=\int_{\mathbb{R}^{3}}\mathcal{H}(\psi,\bar{\psi},\nabla_{x}\psi,\nabla_{x}\bar{\psi})dx,$$
with Hamiltonian density
$$\mathcal{H}=|\nabla_{x}\psi(t,x)|^{2}+\kappa|\psi(t,x)|^{4.}$$
where $\psi$ is a smooth map from $\mathbb{R}_{t}$ to $\mathcal{S}(\mathbb{R}^{3})$. The basic Poisson brackets are
$$\{\psi(x),\bar{\psi}(y)\}=P(x,y),\{\psi(x),\psi(y)\}=0,\{\bar{\psi}(x),\bar{\psi}(y)\}=0,P(x,y)=i\delta(x-y).$$
The equation of motion is
$$i\frac{\partial\psi}{\partial t}=i(H\star_{P}\psi-\psi\star_{P}H)=i\{H,\psi\}=-\triangle_{x}\psi+2\kappa|\psi|^{2}\psi.$$
\end{remark}

\end{document}